\documentclass[twocolumn,prb,showpacs,floatfix,altaffilletter]{revtex4}
\usepackage{graphicx}
\usepackage{bm}
\usepackage{amsmath}
\usepackage{sansmath}
\usepackage{amsfonts}
\usepackage{amssymb}
\usepackage{xcolor}
\newcommand{\alio}{$\alpha$-Li$_2$IrO$_3$\ }
\newcommand{\blio}{$\beta$-Li$_2$IrO$_3$\ }
\newcommand{\glio}{$\gamma$-Li$_2$IrO$_3$\ }
\newcommand*\rfrac[2]{{}^{#1}\!/_{#2}}
\begin{document}
\title{Unconventional Magnetic Order on the Hyperhoneycomb Kitaev Lattice in $\bm{\beta}$-Li$_2$IrO$_3$: \\
Full Solution via Magnetic Resonant X-ray Diffraction}
\author{A.\ Biffin,$^1$ R.\ D. Johnson,$^1$ Sungkyun\ Choi,$^1$ F.\
Freund,$^2$ S.\ Manni,$^2$ \\ A.\ Bombardi,$^3$ P.\ Manuel,$^4$
P.\ Gegenwart$^2$ and R.\ Coldea$^1$}

\affiliation{$^{1}$Clarendon Laboratory, University of Oxford,
Parks Road, Oxford OX1 3PU, U.K.}

\affiliation{$^{2}$EP VI, Center for Electronic Correlations and
Magnetism, Augsburg University, D-86159 Augsburg, Germany}

\affiliation{$^{3}$Diamond Light Source Ltd., Harwell Science and
Innovation Campus, OX11 0DE, U.K.}

\affiliation{$^{4}$ISIS Facility, Rutherford Appleton Laboratory,
Harwell Oxford, OX11 0QX, U.K.}

\pacs{75.25.-j, 75.10.Jm}

\begin{abstract}
The recently-synthesized iridate \blio has been proposed as a
candidate to display novel magnetic behavior stabilized by
frustration effects from bond-dependent, anisotropic interactions
(Kitaev model) on a three-dimensional ``hyperhoneycomb" lattice.
Here we report a combined study using neutron powder diffraction
and magnetic resonant x-ray diffraction to solve the complete
magnetic structure. We find a complex, incommensurate magnetic
order with non-coplanar and counter-rotating Ir moments, which
surprisingly shares many of its features with the related
structural polytype ``stripyhoneycomb" $\gamma$-Li$_2$IrO$_3$,
where dominant Kitaev interactions have been invoked to explain
the stability of the observed magnetic structure. The similarities
of behavior between those two structural polytypes, which have
different global lattice topologies but the same local
connectivity, is strongly suggestive that the same magnetic
interactions and the same underlying mechanism governs the
stability of the magnetic order in both materials, indicating that
both $\beta$- and \glio are strong candidates to realize dominant
Kitaev interactions in a solid state material.
\end{abstract} \maketitle

\section{Introduction}
\label{sec:intro} Materials containing ions with a strong
spin-orbit interaction are attracting much attention as candidates
to display novel electronic states ranging from topological
insulators with protected gapless surface states to quantum spin
liquids with exotic excitations.\cite{witczak} Complex magnetic
behavior can arise when the strong spin-orbit coupling (as found
for heavy transition metal 4$d$ and 5$d$ ions) and crystal field
effects stabilize spin-orbit entangled magnetic moments, which may
interact via anisotropic exchange interactions where the
anisotropy axis depends on the bond orientation. On certain
lattices these interactions may be strongly frustrated, leading
potentially to novel forms of cooperative magnetic order and/or
excitations, not found for magnets of $3d$ ions (where the
spin-orbit coupling is much weaker and the orbital moment is in
general quenched). The magnetic ground state of Ir$^{4+}$ ions in
a cubic IrO$_6$ octahedron is a doublet with a mixed spin-orbital
character,\cite{kim} $J_{\rm eff}=1/2$, and for edge-sharing
IrO$_6$ octahedra it was proposed\cite{jackeli,chaloupka} that the
leading super-exchange is a ferromagnetic Ising coupling between
the Ir magnetic moment components perpendicular to the Ir-O$_2$-Ir
plane. For three-fold coordinated, edge-sharing octahedra this
leads to orthogonal Ir-O$_2$-Ir planes for the three Ir-Ir bonds
and correspondingly orthogonal $\mathsf{x},\mathsf{y},\mathsf{z}$
components coupled along the three Ir-Ir bonds. For a honyecomb
lattice this realizes the Kitaev model,\cite{kitaev} where the
strong frustration effects stabilize an (exactly solvable) quantum
spin liquid state with novel excitations (Majorana fermions and
fluxes).\cite{knolle} In search for realizations of such physics,
$\alpha$-Na$_2$IrO$_3$ [Refs.
\onlinecite{singh,liu,choi,ye,gretarsson}] and
$\alpha$-Li$_2$IrO$_3$ [Refs. \onlinecite{omalley,singh-manni}]
have been actively explored experimentally, however no clear
evidence of novel Kitaev phenomena has been observed.

\begin{figure}[htbp]
\includegraphics[width=0.48\textwidth]{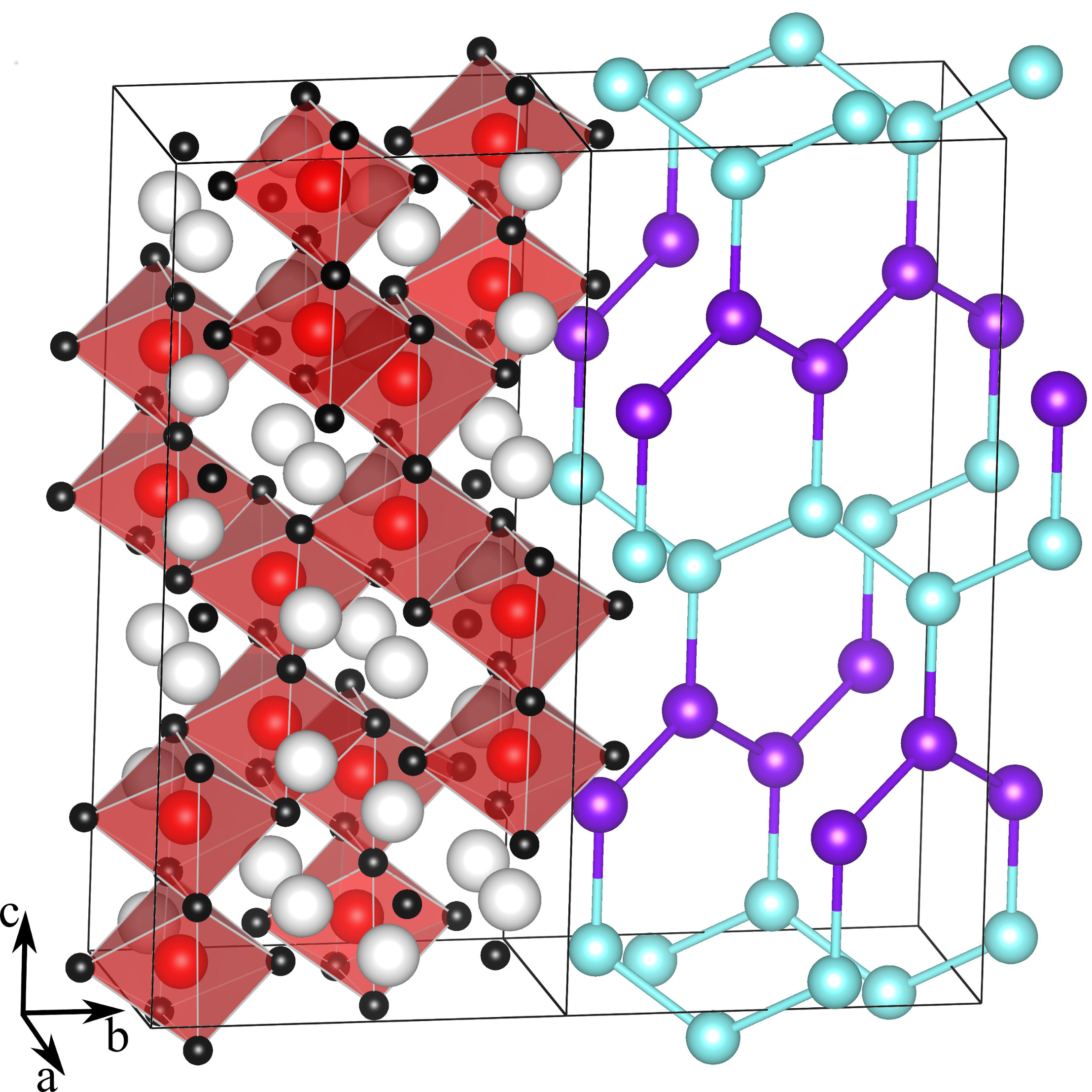}
\caption[]{(color online) Crystal structure of
$\beta$-Li$_2$IrO$_3$. Two unit cells are shown: (left) full
structure with Li (white balls), O (black) and Ir (red) located
inside IrO$_6$ octahedra (shaded polyhedra), (right) Ir ions are
arranged in a ``hyperhoneycomb" structure. Light (cyan) and dark
(purple) colors indicate zig-zag chains stacked along $c$ and
directed alternatingly along $\bm{a}\pm\bm{b}$.}
\label{fig:Fdddstruct}
\end{figure}

Generalization of the Kitaev model to three-dimensional (3D)
lattices have also been shown to have quantum spin liquid ground
states.\cite{mandal,lee,kimchi} Furthermore, non-trivial magnetic
behavior has been predicted when (finite) additional interactions
suppress the spin liquid and stabilize magnetic
order.\cite{lee,lee2} The recently synthesized
``hyperhoneycomb"\cite{takayama} \blio [see Fig.\
\ref{fig:Fdddstruct}] and ``stripyhoneycomb"\cite{modic}
$\gamma$-Li$_2$IrO$_3$, which have the local connectivity of
three-fold coordinated IrO$_6$ octahedra with near-orthogonal
Ir-O$_2$-Ir planes meeting at each Ir site, are prime candidates
to realize 3D Kitaev physics. The availability of single crystal
samples of {\em both} of those two distinct structural polytypes
offers unique opportunities to perform comparative studies between
them and gain insight into the underlying physics. In recent
experiments\cite{cccmpaper} on the $\gamma$-polytype we have
observed an unexpectedly complex, yet highly-symmetric
incommensurate magnetic structure with non-coplanar and
counter-rotating moments, and theoretical calculations showed that
this structure could be stabilized by a spin Hamiltonian with
dominant Kitaev couplings and some additional interactions. Here
we extend our experimental studies to the $\beta$-polytype, where
we find some striking similarities in the magnetic structure. This
shows that the key features of the magnetic order are robust,
independent of the changes in the global lattice topology between
the two distinct structural polytypes that share the same building
blocks. This is strongly suggestive that the same underlying
magnetic interactions govern the stability of the magnetic order
in both structural polytypes, which would be an important
constraint on any theoretical model of the cooperative magnetism
in this family of materials.

The structural polytype \blio was discovered by Takayama {\em et.\
al.} \cite{takayama} and the crystal structure is illustrated in
Fig.\ \ref{fig:Fdddstruct}(left). It has an orthorhombic unit cell
(space group $Fddd$, for full structural details see Appendix
\ref{app:structure}) and the same mass density and same
fundamental building blocks - edge-sharing LiO$_6$ and IrO$_6$
octahedra (the latter shaded in red) as in the layered polytype
\alio (Ref.\ \onlinecite{omalley}), where edge-sharing IrO$_6$
octahedra form planar honeycomb layers separated by hexagonal Li
layers. In the $\beta$-structure [see Fig.\ \ref{fig:Fdddstruct}]
the Ir lattice is three-fold coordinated as in a planar honeycomb,
but now the links of the lattice form a 3D network; this 3D
lattice connectivity has been named a ``hyperhoneycomb".
\cite{takayama} It can be thought of as being made of zig-zag
chains [shaded cyan and purple in Fig.\
\ref{fig:Fdddstruct}(right)] stacked along $c$ and alternating in
direction between the two basal plane diagonals $\bm{a}\pm\bm{b}$
(the planar honeycomb structure is obtained when the zig-zag
iridium chains are {\em not} alternating in orientation, but are
all running parallel to either the $\bm{a}+\bm{b}$, or the
$\bm{a}-\bm{b}$ directions).

Earlier studies\cite{takayama} have shown that \blio is an
insulator. The temperature-dependence of the magnetic
susceptibility parameterized in terms of a Curie-Weiss law gave an
effective magnetic moment $\mu_{\rm eff}=1.61\mu_{\rm B}$,
consistent with localized, $J_{\rm eff}=1/2$ moments at the
iridium sites. Both specific heat and susceptibility data showed
an anomaly near $T_{\rm N}=38$\, K attributed to the onset of
magnetic order. Here we confirm that the low-temperature phase has
magnetic long-range order and we provide a full experimental
magnetic structure solution. Our results give uniquely detailed
information about the correlations that govern the cooperative
magnetism and the relevant magnetic interactions that stabilize
them. We first used neutron diffraction on a powder sample to
confirm the presence of magnetic order at low temperatures and to
obtain candidate magnetic propagation vectors. We then performed
magnetic resonant x-ray diffraction (MRXD) experiments on a
17$\mu$m diameter single crystal at the L$_3$ edge of iridium,
where experiments on other iridates\cite{kim, boseggia, liu} have
reported a strong enhancement of the magnetic scattering
cross-section. We exploit the polarization dependence of the MRXD
intensity (probed via azimuth scans) to deduce that the magnetic
structure has rotating magnetic moments and we determine the plane
of rotation and relative phases between all 16 iridium sites in
the structural unit cell.

The paper is organized as follows: Sec. \ref{sec:npd} presents the
neutron powder diffraction measurements and the analysis by which
we obtain candidate magnetic propagation vectors. Sec.\
\ref{sec:MRXD} presents the single-crystal MRXD measurements,
which observe magnetic Bragg peaks with an incommensurate
propagation vector along the (100) direction. The observed
diffraction pattern is analyzed in terms of magnetic basis vectors
in Sec.\ \ref{sec:basis_vectors} and \ref{sec:selection_rules}.
The azimuth dependence of the diffraction intensity is used to
determine the polarization of all magnetic basis vectors in the
ground state and the relative phase between them (in Sec.\
\ref{sec:azimuths}). Finally the absolute value of the ordered
magnetic moment is extracted from the neutron powder data in Sec.\
\ref{sec:absolute}. The resulting magnetic structure is presented
in Sec.\ \ref{sec:magstructure} and similarities with the magnetic
structure in the $\gamma$-polytype are discussed in Sec.\
\ref{sec:discussion}. Finally, conclusions are summarized in Sec.\
\ref{sec:conclusions}. Further technical details of the analysis
are presented in the Appendices: (A) crystal structure refinement
at low temperature from single-crystal x-ray data, (B)
decomposition of the magnetic structure in terms of its Fourier
components, (C) derivation of the selection rules for magnetic
scattering for the various basis vectors, (D) the azimuth
dependence of the MRXD intensity enabling determination of the
relative phase between basis vectors, and (E) an equivalence
mapping between the magnetic structures in the $\beta$- and
$\gamma$-polytypes of Li$_2$IrO$_3$.

\section{Magnetic Neutron Powder Diffraction}
\label{sec:npd} Neutron powder diffraction measurements to obtain
information about the magnetic propagation vector were performed
using the time-of-flight diffractometer WISH at ISIS. 0.71g of
powder \blio (synthesized as described in Appendix
\ref{app:structure}) was placed in an aluminium can of annular
geometry (to minimize the strong neutron absorption by the Ir
nuclei) and was mounted on the cold finger of a closed cycle
refrigerator. Diffraction patterns were collected at a selection
of temperatures from base (5.6~K) to paramagnetic (70~K), well
above the magnetic ordering transition $T_{\rm N}=38$~K inferred
from thermodynamic measurements.\cite{takayama} The obtained
neutron diffraction pattern did not allow for a full structural
refinement (most likely due to neutron absorption effects in the
presence of iridium). We therefore performed additional x-ray
measurements on a small single crystal piece extracted from the
same powder batch to determine precise internal atomic positions
in the unit cell at a temperature of 100\, K, considered to be
cold enough to be representative of the crystal structure in the
low-temperature limit. The x-ray diffraction was performed on a
Mo-source Oxford Diffraction Supernova x-ray diffractometer under
a N$_2$ gas flow, which gave a temperature of 100\, K at the
sample position. The data confirmed the expected space group and
full structural refinement (for details see Appendix
\ref{app:structure}) gave atomic positions consistent with those
previously reported at room temperature.\cite{takayama}

\begin{figure}[htbp]
\includegraphics[width=0.48\textwidth]{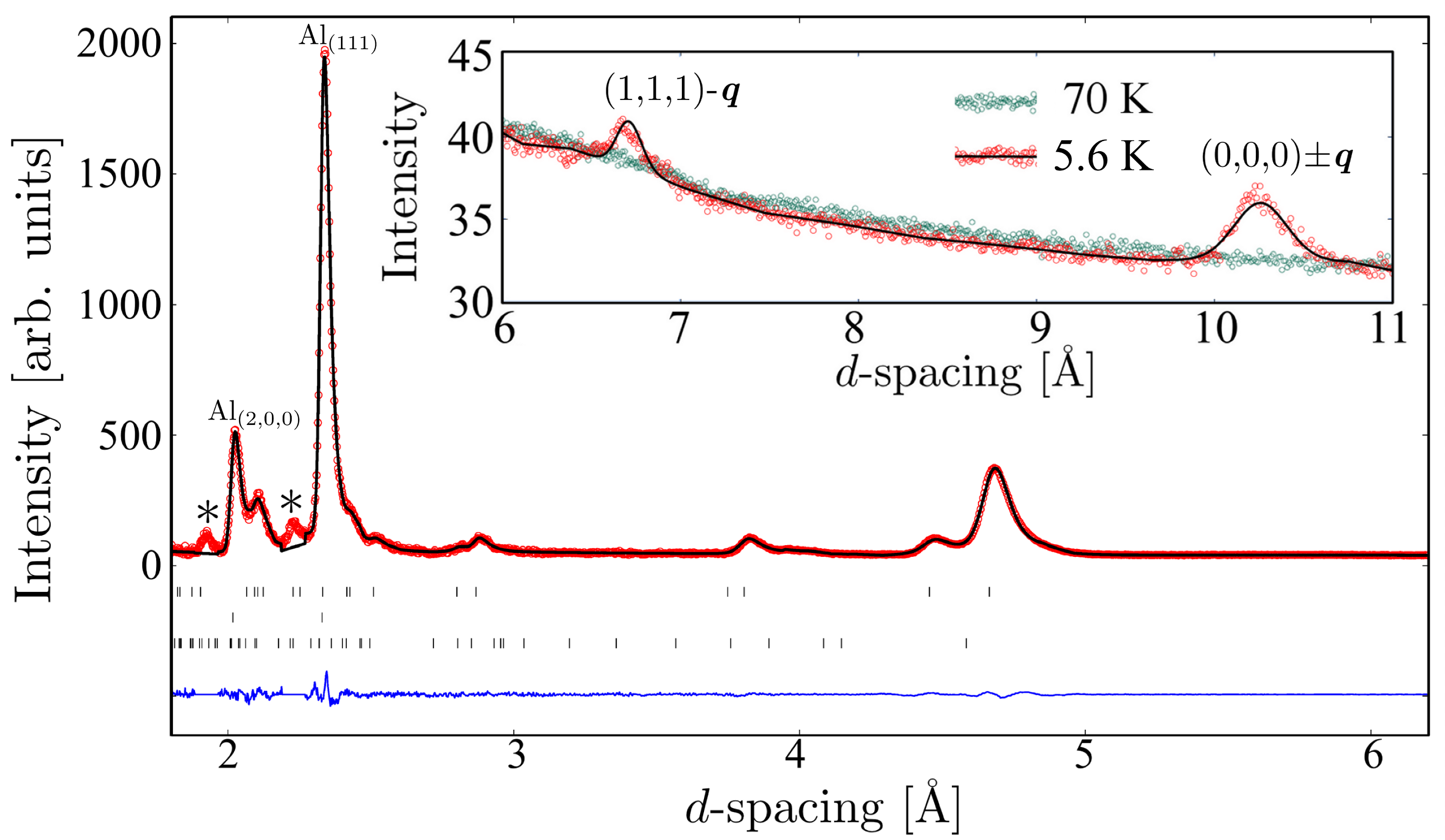}
\caption[]{(color online) Neutron powder diffraction at base
temperature (5.6\, K, red circles) and in the paramagnetic regime
(70\, K, green circles) in the lowest-angle detector bank. Solid
black line shows the fit to the structural and magnetic
contributions as discussed in the text. Positions of structural
\blio peaks, aluminium peaks and magnetic Bragg peaks are marked
below the pattern in the upper, middle and lower rows,
respectively and the blue line underneath represents the
difference between data and fit. Labels ``Al" indicate aluminium
peaks and (*) denote two peaks from an impurity phase identified
as pure Ir, removed from the fit. Inset: zoom into the large
$d$-spacing region showing two magnetic Bragg peaks, labels
indicate the magnetic peaks index (symmetrically equivalent
indexes are omitted).} \label{fig:npd}
\end{figure}

Fig.\ \ref{fig:npd} shows the neutron powder diffraction pattern
in the lowest angle bank of detectors, which covers the region of
large $d$-spacings where magnetic diffraction is expected. In the
main panel the visible diffraction peaks are of structural origin.
\blio peak positions are denoted by the upper row of vertical
marks (below the data) and the two large peaks in the data at
$d$-spacing near 2~\AA\ (labelled ``Al") are the lowest-order
diffraction peaks from the Al can and were fitted as such. The
pattern also showed two peaks labelled (*), originating from a
small impurity phase identified as pure Ir, and those regions in
the data were excluded from the analysis. We performed the
structural fit (using FullProf \cite{fullprof}) to the neutron
data (both paramagnetic and base temperature) with the internal
atomic positions kept fixed, and only the lattice parameters left
free to vary. The structural fit was consistent over all detector
banks - corresponding to a range of scattering angles and
resolutions - and we were also able to consistently fit the
instrument parameters (which provide the conversion from
time-of-flight to $d$-spacing) over all the detector banks tested.
Fig.\ \ref{fig:npd} illustrates the very good agreement in
describing the structural diffraction pattern.

\subsection{Candidate magnetic propagation vectors}
\label{sec:candidate_q} The inset in Fig.\ \ref{fig:npd}
highlights the region of large $d$-spacing, where upon cooling
below $T_{\rm N}$ two new Bragg peaks appear, attributed to the
onset of magnetic ordering. The peak widths are comparable to
those of structural Bragg peaks, implying long-range magnetic
order. With only two magnetic peaks the magnetic structure cannot
be solved uniquely, however the data provides strong constraints
on the magnetic propagation vector, $\bm{q}$. Typically $\bm{q}$
will lie on a point, line or plane of high symmetry in the
Brillouin zone. Using the {\em k-search} tool in FullProf
\cite{fullprof} we systematically searched all such positions for
the $Fddd$ space group in order of decreasing symmetry to find
propagation vectors that could reproduce the magnetic peak
positions observed in the powder neutron data. All commensurate
orderings with wavevectors at the $\Gamma$, $Z$, $Y$, $T$ and $L$
points (where the Miller-Love convention for labelling
high-symmetry Brillouin zone points has been used\cite{miller})
were ruled out by this analysis. Therefore we considered
propagation vectors along lines of high symmetry with general
points denoted by $\Sigma$, $\Delta$, $\Lambda$. Only the first
two were compatible with the neutron data and the obtained
solutions are listed in Table \ref{sym_tab}, along with possible
solutions for other high symmetry directions.

\begin{table}[htbp]
\caption{\label{sym_tab} Candidate magnetic propagation vectors
along high symmetry directions in the Brillouin zone compatible
with the magnetic neutron powder diffraction pattern. The ellipses
denotes the fact that multiple solutions exist for a general
propagation vector in the $ab$-plane.}
\begin{tabular}{c|c}
Position &  $\bm{q}_{\rm fit}$  \\
\hline
$\Sigma$ & (0.57,0,0) \\
$\Delta$ & (0,0.81,0) \\
$J$ & (0.27,0,$\rfrac{1}{2}$) \\
$E$ & (0,0.39,$\rfrac{1}{2}$) \\
$M$ & (0.43,0.53,0) \\
 &  \vdots  \\
\end{tabular}
\end{table}

\section{Magnetic Resonant X-ray Diffraction}
\label{sec:MRXD}

\begin{figure}[htbp]
\includegraphics[width=0.48\textwidth]{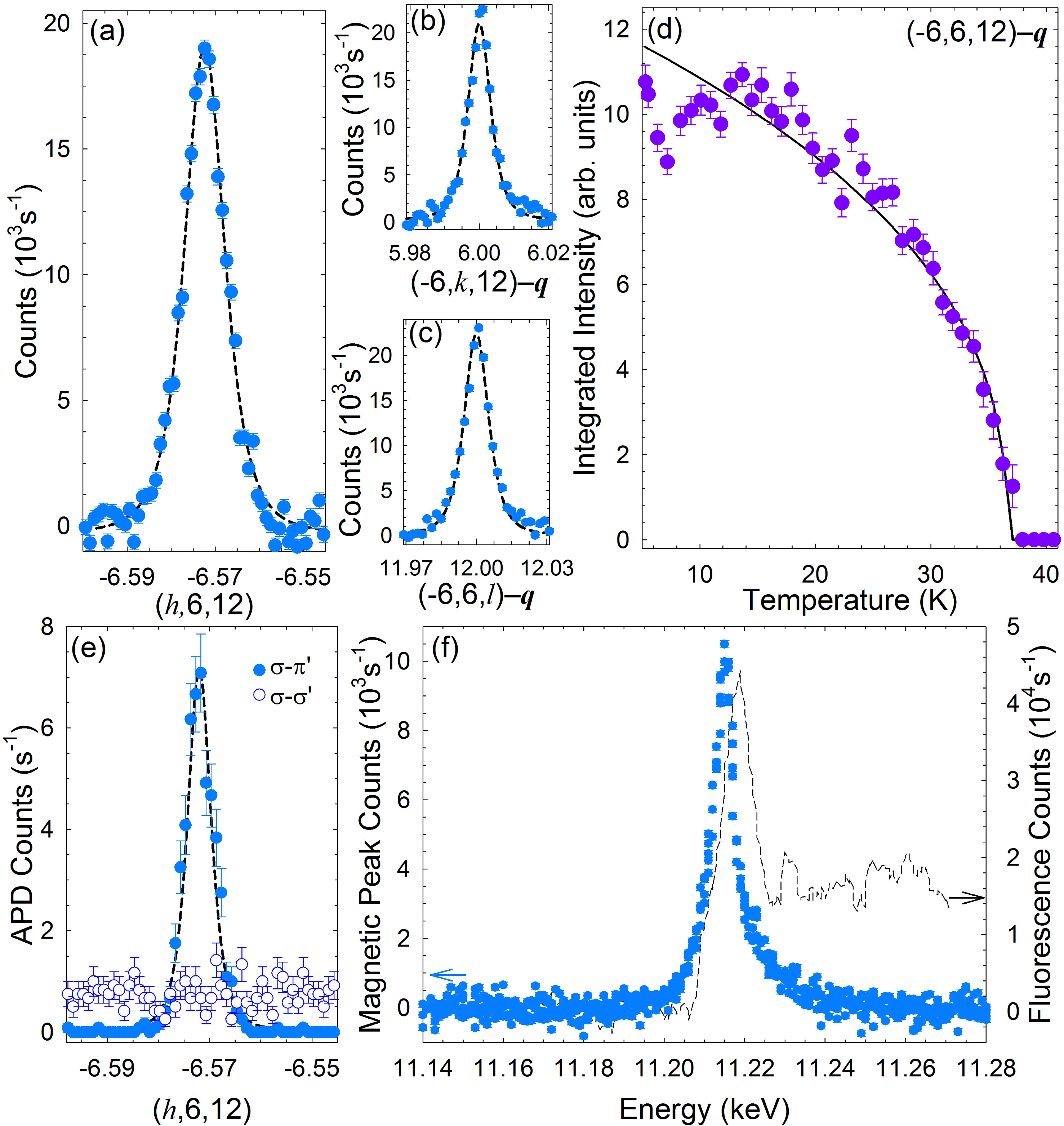}
\caption[]{(color online) Magnetic Bragg peak at
(-6,6,12)$-$$\bm{q}$; (a) Scan shows that the peak occurs at an
incommensurate position along $h$, and is well-centered at integer
(b) $k$, and (c) $l$. (d) Temperature-dependence of the integrated
peak intensity (solid line is a power-law fit).(e) Scans with a
polarizer in the scattered beam show that the peak is only present
in the rotated polarization channel ($\sigma$-$\pi'$) (light blue,
filled) as expected for magnetic scattering. (f) Intensity as a
function of the x-ray energy (filled points) shows that the signal
is resonant on the edge of the fluorescence signal onset (dashed
line), as expected for magnetic scattering.} \label{fig:braggpeak}
\end{figure}

To determine which of the candidate magnetic propagation
wavevectors identified by the neutron data actually occurs, we
have performed a magnetic resonant x-ray diffraction experiment in
reflection geometry at the L$_3$ edge of Ir using the I16 beamline
at Diamond. We used a single crystal of \blio (characterized via
x-ray diffraction as described in Appendix\ \ref{app:structure})
with an orthorhombic morphology with its maximum dimension less
than 17$\mu$m. The sample was placed on a Si (111) disk with the
reciprocal lattice vector (-6,5,7) being approximately surface
normal and it was cooled using a closed-cycle refrigerator with Be
domes. The x-ray energy was tuned to the L$_3$ edge of Iridium
(11.215\, keV) and on cooling below $T_{\rm N}$ new Bragg peaks
appeared at satellite positions of reciprocal lattice points at
$(h,k,l) \pm \bm{q}$, with $h,k,l$ integers and an incommensurate
ordering wavevector $\bm{q}$$=$$(0.57(1),0,0)$. Scans through such
a peak are shown in Fig.\ \ref{fig:braggpeak}a-c). The scans
emphasize the incommensurate position along $h$ and centering at
integer $k$ and $l$ values. In the language of Table\
\ref{sym_tab} this is then a propagation vector along the $\Sigma$
line of symmetry, and the magnitude found from the x-ray data is
consistent with the positions of peaks in the neutron data (first
line in Table\ \ref{sym_tab}). Lorentzian squared fits to the peak
shape profiles (dashed lines in Figs.\ \ref{fig:braggpeak}a-c),e)
gave widths comparable to those measured for the nearby structural
Bragg peak (-6,6,12), indicating that the magnetic order is
long-range in all three directions. To estimate a lower bound on
the correlation lengths we use the inverse of the peak
half-widths, which gives values in excess of 100, 200 and
300~\AA~along the $a$, $b$ and $c$ axes, respectively (more than
$\approx$20 unit cells in each direction), and in reality the
correlation lengths will be much larger as the above estimate did
not include the peak broadening due to the finite instrumental
angular resolution.

Several independent tests can be performed to confirm the magnetic
origin of the observed diffraction peaks. Firstly, in Fig.\
\ref{fig:braggpeak}(d) the integrated intensity of
(-6,6,12)-$\bm{q}$ as a function of temperature (filled circles)
shows a typical order-parameter behavior. A power-law fit (solid
line) gives an onset temperature of $T_{\rm fit}=36.7$~K, which is
very close to the transition temperature deduced from specific
heat measurements.\cite{takayama} We attribute the apparent small
offset of $\Delta T$$\simeq$1.3~K to the fact that the sensor
where the temperature was measured in the x-ray experiments was at
the bottom of the sample support and the actual sample temperature
may have been higher by $\Delta T$ due to local heating effects
from the very intense synchrotron beam, a common occurrence in
resonant x-ray experiments.

Further evidence of the magnetic origin of the incommensurate
diffraction peaks can be obtained via polarization analysis of the
scattered beam. For a scattering experiment with incident beam
polarization $\bm{\hat{\epsilon}}$ normal to the
$(\bm{k},\bm{k'})$ scattering plane ($\sigma$-polarized), magnetic
scattering is expected to rotate the polarization by 90$^{\circ}$
compared to the incident beam direction (scattering in the
$\sigma$-$\pi'$ channel, see diagram in Fig.\ \ref{fig:basisazi}a)
(inset)), whereas conventional (charge) scattering would leave the
polarization direction unchanged (scattering in the
$\sigma$-$\sigma'$ channel).\cite{hill} In the experiments
polarization analysis was achieved via the 3$^{\rm rd}$ harmonic
reflection from the (1,1,1) planes of an Au crystal, the
$d$-spacing of 0.785~\AA~being ideally suited for polarization
analysis at the L$_3$ edge of Ir (giving perfect filtering at
11.16~keV, see Ref. \onlinecite{detlefs}). Fig.\
\ref{fig:braggpeak}(e) shows an $h$-scan across the incommensurate
peak performed with polarizer parallel (open symbols,
$\sigma$-$\sigma'$) and perpendicular (filled symbols,
$\sigma$-$\pi'$) to the original incident beam polarization. The
signal appears only in the rotated polarization channel, as
expected for magnetic scattering. Note that the higher background
in the $\sigma$-$\sigma'$ channel is due to incoherent x-ray
scattering, which is largely filtered out by the polarization
analysis in the $\sigma$-$\pi'$ channel revealing a pure magnetic
signal.

A third test is the dependence on the x-ray energy as the magnetic
scattering is only expected at resonance. Indeed, the
incommensurate peak intensity has a characteristic resonant
response as shown in Fig.\ \ref{fig:braggpeak}(f) (filled blue
symbols) with a maximum at the resonance energy of 11.215~keV in
agreement with previous resonant studies on other
iridates,\cite{liu,boseggia} and coincides with the edge of the
fluorescence signal from the sample (black, dotted line).

\subsection{Magnetic Basis Vectors}
\label{sec:basis_vectors} The observation of incommensurate
magnetic Bragg peaks indicates a magnetic structure with rotating
moments or a spin-density wave, both with a wavelength that is
incommensurate with the underlying crystal lattice periodicity. In
order to distinguish between those two scenarios, and to determine
fully the orientations of the magnetic moments at all magnetic
sites in the unit cell, we consider below symmetry-allowed basis
vectors for magnetic structures compatible with the observed
propagation vector $\bm{q}$. This provides a natural framework to
directly link diffraction data with a magnetic structure model and
allows one to develop a systematic strategy in experiments to
determine the magnetic structure completely.

The magnetic ions, Ir$^{4+}$, occupy a single crystallographic
site with ions at $(1/8,1/8,z)$ [$z=0.70845(7)\simeq$ 5/8+1/12]
and symmetry equivalent positions, giving four Ir ions in the
primitive unit cell (labelled 1-4 in Fig.\ \ref{fig:magstruct} and
with positions listed explicitly in Table\ \ref{tab:magstruct} in
Appendix \ref{app:magnetic}). Taking into account that at each
site the magnetic moment could have components along the $x$, $y$
and $z$-axes (along the orthorhombic $a$, $b$, $c$-axes), gives a
12-component representation of the magnetic structure. Its
irreducible representations and associated magnetic basis vectors
(obtained using the $BasIrreps$ tool, part of the FullProf
suite\cite{fullprof}) are listed in Table\ \ref{irrep_tab}.

\begin{table}[h!]
\caption{\label{irrep_tab} Irreducible representations and basis
vectors for a magnetic structure with propagation vector
$\bm{q}$$=$$(q,0,0)$.}
\begin{tabular}{c|c}
Irreducible & Basis Vectors \\
Representation & \\
\hline
$\Gamma_1$ & $F_x, G_y, A_z$\\
$\Gamma_2$ & $C_x, A_y, G_z$\\
$\Gamma_3$ & $G_x, F_y, C_z$\\
$\Gamma_4$ & $A_x, C_y, F_z$\\
\end{tabular}
\end{table}

The basis vectors contain symmetry-imposed relations between the
Fourier components of the magnetic structure between the four
sites in the primitive cell. The types of basis vectors that can
occur are:
\begin{equation} F=\left[
\begin{array}{c}
1\\
1\\
\delta \\
\delta
\end{array} \right], C= \left[ \begin{array}{c}
~~1\\
~~1\\
-\delta\\
-\delta
\end{array} \right], A= \left[ \begin{array}{c}
~~1\\
-1\\
-\delta\\
~~\delta
\end{array} \right], G= \left[ \begin{array}{c}
~~1\\
-1\\
~~\delta\\
-\delta
\end{array} \right],
\label{eq:FCAG}
\end{equation}
where for each vector the four values are relative phase factors
between the Fourier components at the four sites.
$\delta=e^{-i\bm{q}\cdot(\bm{r}_3-\bm{r}_1)}=e^{-i\pi q/2}$ is a
displacement phase factor that takes into account the fact that
sites 3 and 4 are displaced relative to sites 1 and 2 by $a/4$ in
the direction of the propagation vector $\bm{q}$. In an $F$ basis
vector the Fourier components at the 4 sites are related by
$\bm{M}_{{\bm q},1}=\bm{M}_{{\bm q},2}=\delta^{-1}\bm{M}_{{\bm
q},3}=\delta^{-1}\bm{M}_{{\bm q},4}$, i.e. there are no additional
phase factors between the four sites apart from the natural
displacement phase factor, and in the limit $q\rightarrow0$
($\delta \rightarrow 1$) one recovers ferromagnetic order. For the
basis vectors $C$, $A$ and $G$ two out of the four sites have a
change in sign in addition to the normal displacement phase
factor, and in the limit $q\rightarrow0$ one finds an
antiferromagnetic alignment between the four sites. To determine
the full magnetic structure, we first identify which basis vectors
are present, determine their polarization (i.e. along $x$, $y$ or
$z$) and then find the relative phase between them, as follows.

\subsection{Selection Rules for Magnetic Scattering}
\label{sec:selection_rules} Each magnetic basis vector contains a
strict, symmetry imposed phase relation between moments on all Ir
sites in the primitive unit cell. As such, one can derive
selection rules for non-zero magnetic diffraction intensity, i.e.,
each basis vector will only contribute to magnetic satellite
reflections of certain reciprocal lattice points and will have
zero structure factor for others. It follows that simply the
presence or absence of magnetic Bragg peaks at certain positions
can already identify which basis vectors are present.

Explicitly, the structure factor for a magnetic superlattice
reflection at $\bm{Q}=(h,k,l) \pm \bm{q}$ is
\begin{equation}
\bm{\mathcal{F}}(\bm{Q})=
\bm{\mathcal{F}}((h,k,l){\pm}\bm{q})=f_{F}~\sum_{n}\bm{M}_{\pm
\bm{q},n}e^{i\bm{Q \cdot r}_n}, \label{eq:structfact}
\end{equation}
where the pre-factor
$f_F=1+e^{i\pi(h+k)}+e^{i\pi(k+l)}+e^{i\pi(l+h)}$ is due to the
$F$-centering of the orthorhombic structural unit cell. The sum
extends over all sites in the primitive cell ($n$=1$-$4) and
$\bm{M}_{\pm \bm{q},n}$ are the Fourier components of the magnetic
moments at site $n$ with position in the unit cell $\bm{r}_n$. For
an $F$-basis vector we obtain the structure factor
\begin{equation}
\bm{\mathcal{F}}^F(\bm{Q})= \left\{
\begin{array}{l}
16\bm{M}_{\pm \bm{q},1}\cos\frac{\pi l}{6} e^{i \xi_{ \pm}}, ~~ h,k,l ~ \text{all even and}\\
\quad \quad \quad \quad \quad \quad h+k+l=4p, ~ p ~ \text{integer} \\
16\bm{M}_{\pm \bm{q},1}\cos \frac{\pi l}{6} \cos \frac{\pi (h+l-k)}{4} e^{i \chi_{ \pm}}, \\
\quad \quad \quad \quad \quad \quad   h,k,l  ~\text{all odd}\\
0, \quad \quad \quad \quad \quad ~ \text{otherwise},
\end{array} \right.
\label{eq:structF}
\end{equation}
where $\xi_{ \pm}=\pi(h+k-3l ~\pm q)/4$ and $\chi_{ \pm}=\pi(2 h-2
l~ \pm q)/4$ and to obtain a closed-form analytic expression we
used the ``ideal" iridium position $z$=5/8+1/12. From
(\ref{eq:structF}) we conclude that an $F$ basis vector will
contribute to the intensity of $\pm\bm{q}$ satellites of
reciprocal lattice points ($h,k,l$) with the selection rule
$h,k,l$ all odd, or all even with $h+k+l=4p$, $p$ integer. This is
the same selection rule as for structural Bragg scattering from
the iridium ion sublattice at the ($h,k,l$) position, as expected
since in the limit $q\rightarrow 0$ the $F$-type magnetic basis
vectors recover ferromagnetic order, and this has the same
selection rule as structural scattering. Structure factor
expressions for the other basis vectors are given in Appendix
\ref{app:magfactors}. The resulting selection rule for $A$ is same
as for $F$, but with no contribution from $A$ for $l=6m$, $m$
integer. The selection rule for $C$ and $G$ is $h,k,l$ all odd, or
all even and $h+k+l=4p+2$, $p$ integer with no contribution from
$G$ if $l=6m$, $m$ integer.

\begin{figure}[htbp]
\includegraphics[width=0.48\textwidth]{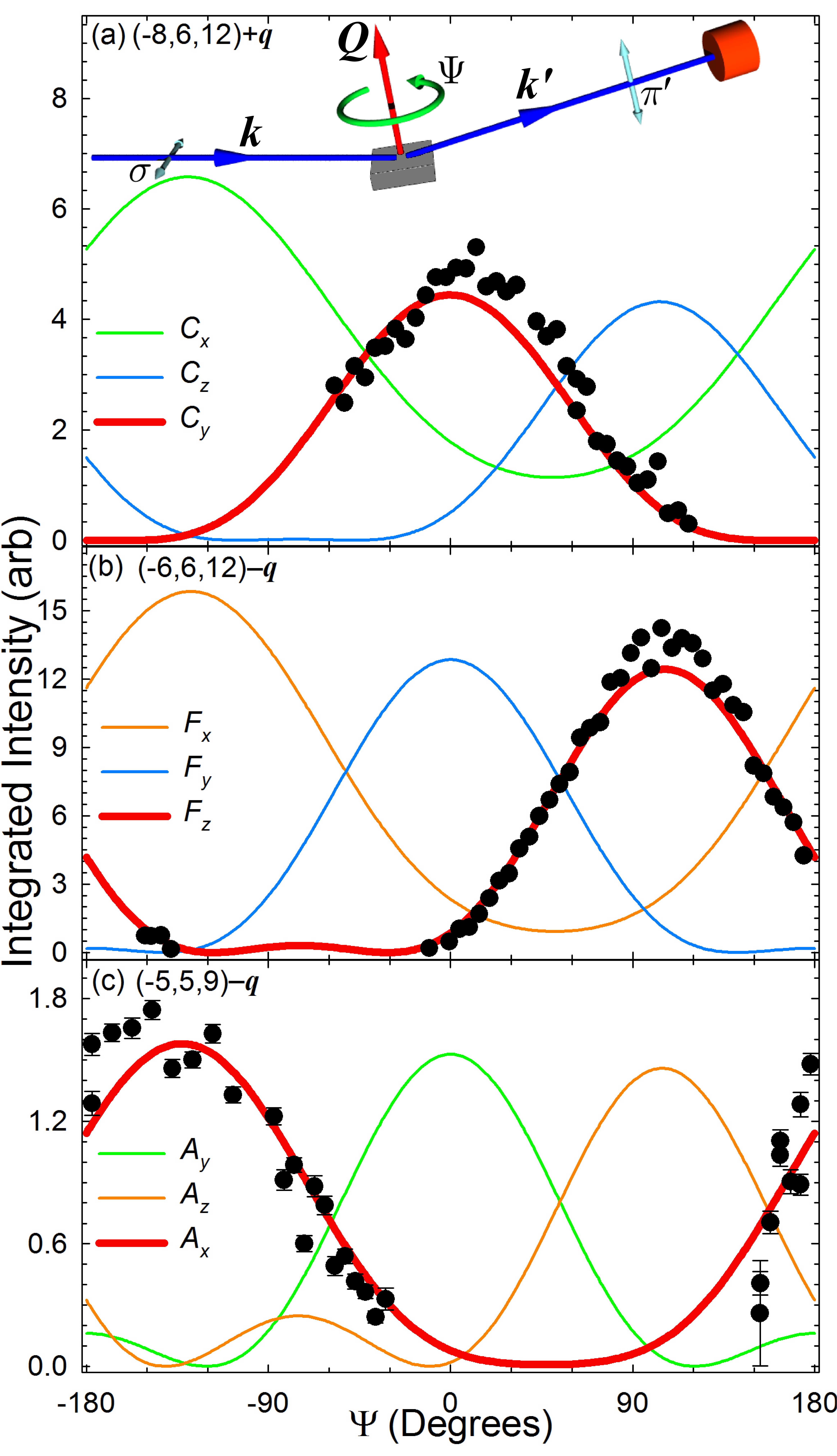}
\caption[]{(color online) Intensity as a function of azimuth for
three magnetic Bragg peaks a) $C_y$, b) $F_z$ and c) $A_x$. Top
diagram illustrates the scattering geometry. Data points (filled
circles) are integrated peak intensities (in sample rocking curve
scans) corrected for absorption and Lorentz factor. Thick (red)
lines show fits that include all contributions to the magnetic
scattering intensity\cite{magnetix} for the magnetic structure
model $\Gamma_4$ depicted in Fig.\ \ref{fig:magstruct}. Orange,
green and blue lines denote other models ($\Gamma_{1,2,3}$,
respectively), which can be easily ruled out. The azimuth origin
$\Psi=0$ corresponds to the case when the ($010$) axis is in the
scattering plane making the smallest angle with the incident beam
direction $\bm{\hat{k}}$.\label{fig:basisazi}}
\end{figure}

The experimental geometry constrained the search for magnetic
satellite reflections to regions of reciprocal space in the
vicinity of the surface normal reflection, $\approx$(-6,5,7).
Using (\ref{eq:structF}) and similar expressions for $C,G$ and $A$
(see Appendix \ref{app:magfactors}) we ascertained 3 positions of
interest for identifying which basis vectors contribute to the
magnetic structure, namely (-6,6,12)$\pm \bm{q}$ (pure $F$),
(-8,6,12)$\pm \bm{q}$ (pure $C$) and (-5,5,9)$\pm\bm{q}$ (mixed
$A$ and $G$) and found magnetic peaks at all these positions.
Therefore we deduce that the structure contains $F$, $C$ and one
(or both) of $A$ and $G$. To determine the direction of the
magnetic moment components for each of those basis vectors we make
use of the polarization dependence of the magnetic x-ray
diffraction intensity, as follows.

\subsection{Azimuth Scans to Determine the Moment's Direction and Relative Phases Between Basis Vectors}
\label{sec:azimuths} In the dipolar approximation the magnetic
x-ray diffraction intensity at resonance is proportional to
\begin{equation}
L(\theta) \mathcal{A} \left|(\bm{\hat{\epsilon'}} \times
\bm{\hat{\epsilon}}) \cdot\bm{\mathcal{F}}(\bm{Q})\right|^2,
\nonumber 
\end{equation}
where $L(\theta$) is the Lorentz factor at the scattering angle
$2\theta$, $\mathcal{A}$ is an absorption correction dependent
upon the experimental geometry, $\bm{\mathcal{F}}(\bm{Q})$ is the
magnetic structure factor vector given in (\ref{eq:structfact}),
and $\bm{\hat{\epsilon'}}$ and $\bm{\hat{\epsilon}}$ are unit
vectors along the polarization of the electric field component of
the scattered and incident x-ray beams, respectively. \cite{hill}
For a $\sigma$-polarized incident beam magnetic resonant
scattering occurs only in the $\sigma$-$\pi'$ channel [see diagram
in Fig.\ \ref{fig:basisazi}a) inset], meaning that the product of
the electric field polarization vectors is along the scattered
beam direction, i.e. $\bm{\hat{\epsilon'}} \times
\bm{\hat{\epsilon}}=\bm{\hat{k'}}$, so only the component of the
structure factor vector along the scattered beam direction,
$\mathcal{F}_{\parallel}=\bm{\mathcal{F}}\cdot\bm{\hat{k'}}$,
contributes to the magnetic intensity. By keeping the instrument
in the scattering condition and rotating the sample around the
scattering wavevector, $\bm{Q}=\bm{k'}-\bm{k}$, the projection
$\mathcal{F}_{\parallel}$ of the structure factor vector onto the
(fixed) direction $\bm{\hat{k'}}$ varies depending on the azimuth
angle $\Psi$, with maximum magnetic intensity when the magnetic
moments that give rise to the scattering make the smallest angle
with $\bm{\hat{k'}}$ and zero intensity when they are
perpendicular.

Azimuth scans to test the polarization dependence are shown in
Figs.\ \ref{fig:basisazi}a-c) where the data points are obtained
by integrating the magnetic peak intensity in sample rocking curve
scans for each value of the azimuth, $\Psi$. As discussed earlier,
the selection rules identify the signal in panel a) at
(-8,6,12)$+\bm{q}$ to be of $C$-character and the observed maximum
intensity at $\Psi=0$ uniquely identifies the polarization to be
along the $y$-axis (red thick line) as at $\Psi=0$ the $y$-axis is
in the scattering plane and makes the smallest angle with the
$\bm{\hat{k'}}$ direction and makes a larger angle with
$\bm{\hat{k'}}$ when the azimuth is displaced away from 0 in both
directions; green and blue curves show the expected intensity
dependence for other polarizations that can be clearly ruled out.
Similarly, in panel b) the peak at (-6,6,12)$-\bm{q}$ is
identified as being of $F$ character and polarized along $z$
(thick red line). Panel c) shows data at (-5,5,9)$-\bm{q}$,
clearly of $x$-polarization (thick red line). The structure factor
is compatible with both an $A$ or $G$ character and we will show
later that only a pure $A$ character can describe this and other
azimuth scans quantitatively, so we identify this peak as coming
from an $A_x$ basis vector.

\begin{figure}[htbp]
\includegraphics[width=0.48\textwidth]{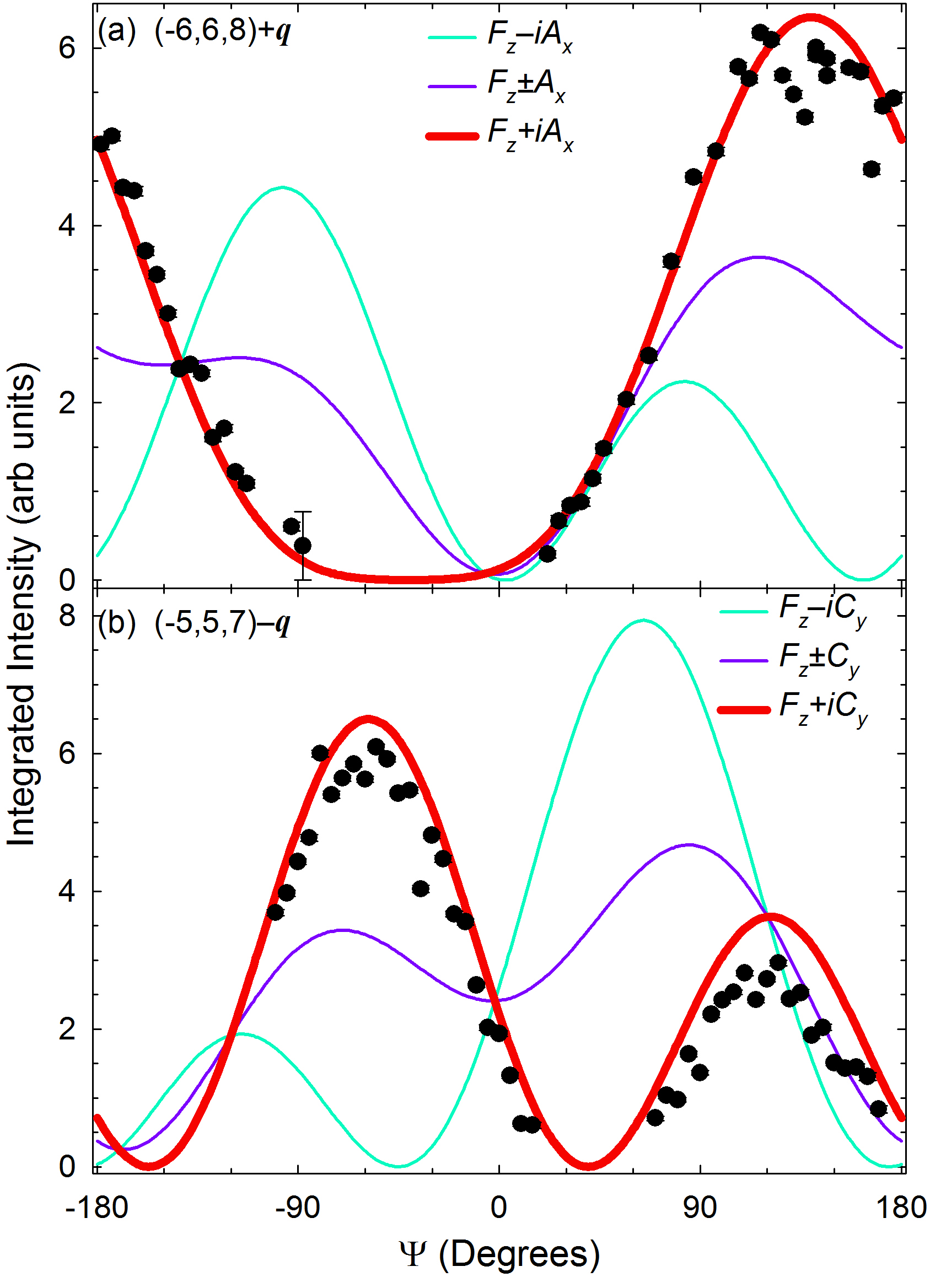}
\caption[]{(color online) Intensity as a function of azimuth for
``mixed" magnetic Bragg peaks (black circles) corrected for
absorption and Lorentz factor, compared to calculations (red thick
line) for the magnetic structure model in Fig.\
\ref{fig:magstruct}, as well as calculations for other possible
models (a) (-6,6,8)$+\bm{q}$ is a mixed $F:A$ peak with
contributions in the ratio $1:3$, (b) (-5,5,7)$-\bm{q}$ is a mixed
$F:C:A$ peak with contributions in the ratio $3:3:1$. Note that
the data clearly distinguishes between different relative phases
of basis vectors (cyan and purple).} \label{fig:phaseazi}
\end{figure}

The next step in solving the structure is to determine the
relative phase between the $A_x$, $C_y$ and $F_z$ basis vectors. A
phase of 0 or $\pi$ between basis vectors with orthogonal
components, say along $A_x$ and $F_z$, means that the moments vary
sinusoidally along a direction in the $xz$ plane, whereas a phase
offset of $\pi/2$ means that the moments rotate in the $xz$ plane.
The relative phase between two basis vectors can be determined
from an azimuth scan at a magnetic Bragg peak position where both
basis vectors contribute as the intensity is the sum of the
intensities due to the two separate basis vectors plus an
additional cross-term that is sensitive to the phase offset (for
more details see Appendix \ref{app:magfactors}).

Fig.\ \ref{fig:phaseazi}(a) shows the azimuth scan at
(-6,6,8)$+\bm{q}$ where structure-factor calculations show that
both $F$ and $A$ basis vectors contribute in the ratio 1:3. The
intensity dependence uniquely identifies it as originating from
the particular basis vector combination $F_z+ iA_x$, i.e. the $x$
components are $\pi/2$ out-of-phase with respect to the
$z$-components, meaning that the magnetic moments rotate in the
$xz$ plane (in a particular sense on each site, note that the
opposite sense of rotations encoded in the basis vector $F_z-iA_x$
is not consistent with the data - cyan line). The data clearly
rules out the case when the moments are not rotating in the $xz$
plane, but are varying sinusoidally along a line in the $xz$ plane
($F_z \pm A_x$, purple line), a model with no $A_x$ component is
also ruled out (not shown). This analysis shows that an $A_x$
basis vector must exist in the ground state and that it is $\pi/2$
out-of-phase with respect to $F_z$. Similarly, to determine the
phase offset between the $F_z$ and $C_y$ basis vectors we measured
the azimuth scan at the magnetic reflection (-5,5,7)$-\bm{q}$ of
mixed $F$, $C$ and $A$ contributions in the ratio 3:3:1, shown in
Fig.\ \ref{fig:phaseazi}(b), and only a model with the basis
vector contribution $F_z+iC_y$ (thick red line) matches the data,
with other combinations (blue and green curves) clearly ruled out
(note that those calculations also include the $iA_x$ contribution
as already determined).

Having established all the basis vectors that are present, their
polarization, and their relative phases, the relative magnitudes
are obtained by quantitatively fitting the azimuth intensity
dependencies. Three free parameters were fit; the relative ratios
$M_x/M_z$ and $M_y/M_z$, and a {\em single} overall intensity
scale factor for all the data. The simultaneous fit of this
magnetic structure model to the data in Figs.\
\ref{fig:basisazi}a-c) and Fig.\ \ref{fig:phaseazi}a-b) (shown by
the thick red solid lines) gave the moment amplitudes in the
relative ratio $M_x:M_y:M_z = 0.45(1):0.65(1):1$. Note that the
basis vector combination $(iA_x,iC_y,F_z)$ with the above moment
amplitudes quantitatively explains the mixed $F_z,A_x$ peak in
Fig.\ \ref{fig:phaseazi}a), and the {\em same} magnitude $A_x$
basis vector also quantitatively accounts for the azimuth
dependence in Fig.\ \ref{fig:basisazi}c), further confirming the
identification of the signal in this magnetic Bragg peak as coming
from a pure $A_x$ basis vector (if a $G_x$ basis vector is also
present and contributing to the signal in Fig.\
\ref{fig:basisazi}c), its magnitude is very small, below the
accuracy of the present experiments). Given this we conclude that
the magnetic structure is described by the basis vector
combination $(iA_x,iC_y, F_z)$, which corresponds to a single
irreducible representation, $\Gamma_4$ in Table\ \ref{irrep_tab}.

\subsection{Absolute Value of the Ordered Moment}
\label{sec:absolute} The only remaining parameter still to be
determined is the absolute magnitude of the magnetic moments. This
is difficult to extract reliably from the magnetic resonant x-ray
diffraction data as it requires accurate determination of scale
factors between the magnetic and structural peaks (the latter
being of order $10^5-10^6$ more intense). However, in neutron
scattering experiments the magnetic and structural peaks are of
comparable magnitudes and one can obtain a reliable determination
of the relative scale factor. To this aim we return to the powder
neutron data in Fig.\ \ref{fig:npd} and simultaneously fit three
contributions to the data: structural peaks of the sample
calculated using the crystal structure with fixed internal
parameters (as deduced from single crystal x-ray diffraction),
aluminium structural peaks (from the sample can), and magnetic
peaks calculated using the full magnetic structure model deduced
from the resonant x-ray experiments. The fit is plotted as a solid
black line in Fig.\ \ref{fig:npd} and shows excellent agreement
with the data. The resulting magnetic moment is of magnitude
0.47(1)$\mu_{\rm B}$ when aligned along the $c$-axis, and slightly
reduced by $\sim$20\% when in the $ab$ plane. The propagation
vector was also fitted and we find $\bm{q}$$=$$(0.5768(3),0,0)$,
consistent with the x-ray measurements.

\subsection{Magnetic Structure} \label{sec:magstructure} 
\begin{figure}[htbp]
\includegraphics[width=0.43\textwidth]{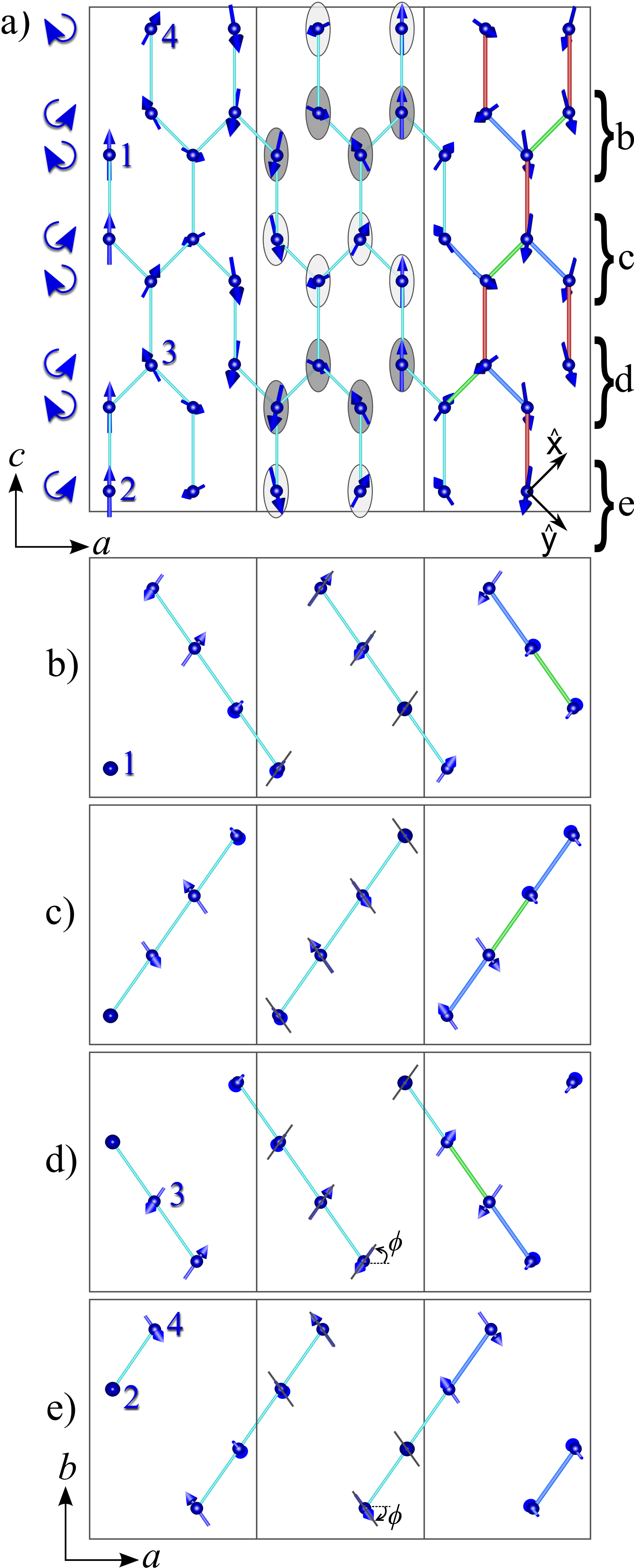}
\caption[]{(color online) a) Projection of the magnetic structure
on the $ac$ plane showing the sites 1-4 of the primitive cell.
Left curly arrows indicate counter-rotation of moments between
consecutive sites along $c$. In unit cell 2 light and dark shaded
elliptical envelopes indicate an alternating tilt of the plane of
moments' rotation away from the $ac$ face. In unit cell 3 the
color of bonds shows the anisotropy axis of the exchange in a
Kitaev model\cite{lee} (ferromagnetic Ising exchange for each
bond, but with a different Ising axis
$\mathsf{x},\mathsf{y},\mathsf{z}$ for the blue/green/red bonds).
Right-hand labels (b)-(e) indicate where slices through the
magnetic structure are taken at different heights in the unit cell
and projected onto the $ab$ plane to illustrate the direction of
the zig-zag chains and the alternating tilt of the plane of
rotation away from the $ac$ plane by $\pm\phi$ between adjacent
zig-zag chains stacked along $c$.} \label{fig:magstruct}
\end{figure}

\begin{figure}[htbp]
\vspace{1mm}
\includegraphics[width=0.365\textwidth]{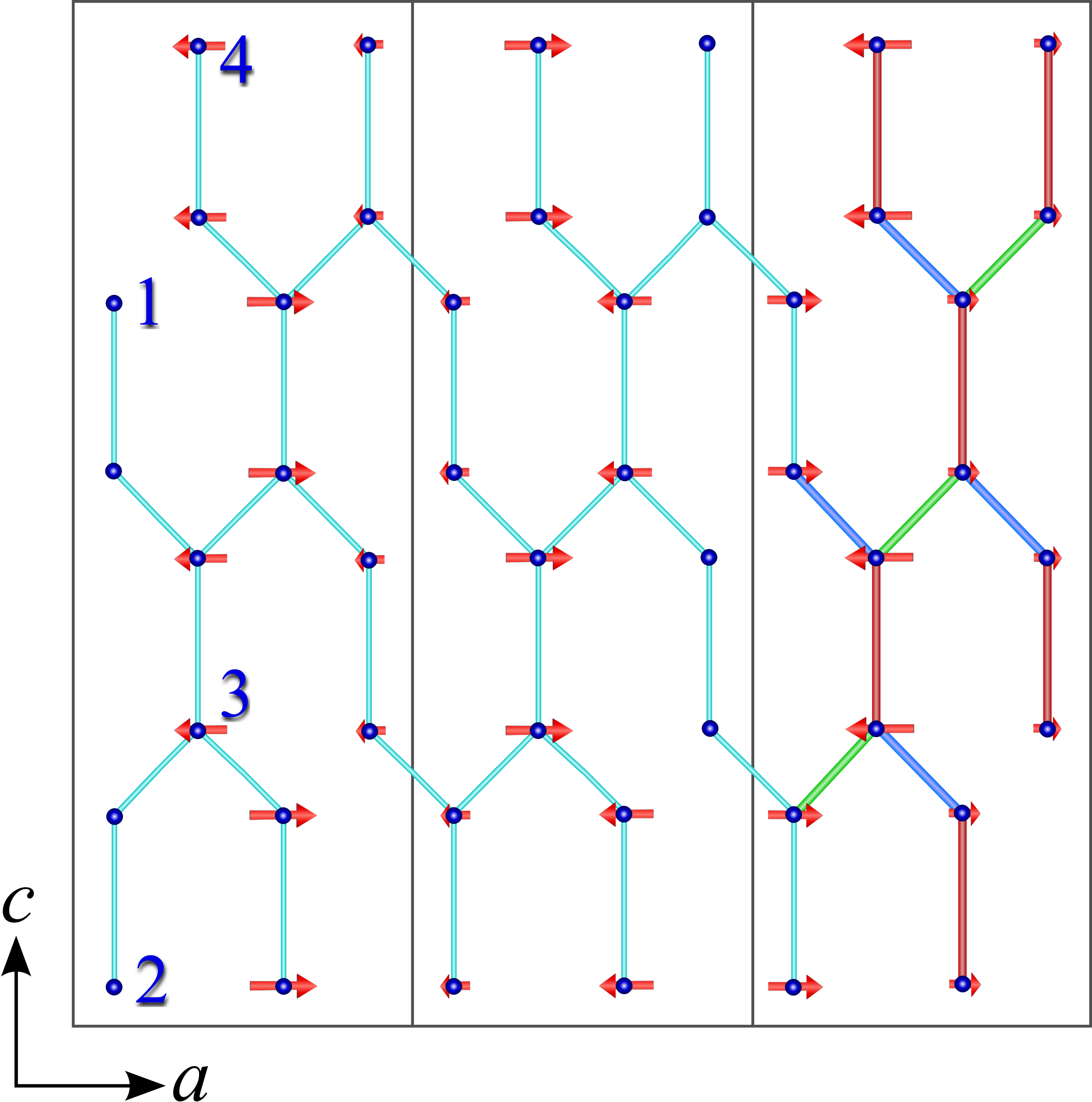}
\caption[]{(Color online) Projection of the iridium lattice on the
$ac$ plane showing the ordering of the magnetic moment components
along the $b$-axis only. These are shown rotated (for ease of
visualization) from the $b$ to the $a$-axis and indicated by
horizontal red arrows (length of arrow indicates magnitude of
$M_y({\bm{r}})$ at each site. Note the ferromagnetic alignment
between the two sites of each vertical ($c$-axis)
bond.}\label{fig:magstruct_My}
\end{figure}

The
obtained magnetic structure projected onto the $ac$ plane is shown
in Fig.\ \ref{fig:magstruct}a). The magnetic moments are
counter-rotating between all nearest-neighbor sites and the plane
of rotation alternates between the two sites of every vertical
($c$-axis) bond as illustrated by the pattern of light and dark
shaded elliptical envelopes. The planes of rotation are obtained
from the $ac$ plane by rotation around the $c$-axis by an angle
$\pm\phi$, with $\phi={\rm
tan}^{-1}\frac{M_y}{M_x}=55(1)^{\circ}$. Each zig-zag chain has
the top and bottom sites counter-rotating in one and the same
plane, which then alternates between consecutive zig-zag chains
vertically-linked along the $c$-axis. The zig-zag chains are
directed alternatingly along the $\bm{a}\pm\bm{b}$ directions and
are made up of iridium sites of type (2,4) and (1,3),
respectively, with vertical ($c$-axis) bonds coupling sites of
types 1-2 and 3-4. The magnetic order pattern is such that for
every vertical bond the spin components along the $y$-axis are
ferromagnetically-aligned (see Fig.\ \ref{fig:magstruct_My}), in
accordance with the basis vector $C_y$, which has equal Fourier
components at sites 1-2 and 3-4, respectively [see eq.\
(\ref{eq:FCAG})].

We note that the magnetic moment rotation at each site defines an
elliptical envelope, distorted from circular with the moment
slightly smaller when in the $ab$ plane compared to when along the
$c$-axis, $\sqrt{M_x^2+M_y^2}/M_z=0.80(1)$. This effect may be due
to a larger $g$-factor along the $c$-axis compared to the
$\bm{\hat{a}}\cos \phi \pm \bm{\hat{b}}\sin \phi $ directions in
the $ab$ plane. We note that single-crystal susceptibility
data\cite{modic} in the related polytype \glio do provide evidence
for the presence of $g$-tensor anisotropy for the iridium moments
(in the high-temperature limit the susceptibility, $\chi$, is
expected to be proportional to the squared $g$-factor along the
applied field direction, and experiments observe $\chi_c$ larger
than both $\chi_a$ and $\chi_b$, implying an anisotropic
$g$-tensor). The very similar local environment around the iridium
sites in \blio suggests that an anisotropic $g$-tensor is also
likely here. Another possibility is that zero-point quantum
fluctuations may prefer a non-fixed length ordered moment between
sites, with a larger ordered moment along the $c$-axis, which is
the only direction that is common to all planes of rotation.
Evidence for the presence of zero-point quantum fluctuations in
the ground state is provided by the fact that the absolute
magnitude of the ordered moment (0.47(1) $\mu_B$) is significantly
reduced from what is believed to be the available full-moment
value (estimated at $g\mu_B J_{\rm eff} \simeq 1\mu_B$ assuming $g
\simeq 2$), so a structure with small modulations on an already
significantly reduced ordered moment could be compatible with the
experimental results.

\section{Discussion}
\label{sec:discussion} It is interesting to note that the obtained
magnetic structure has striking similarities with the magnetic
structure in the related polytype $\gamma$-Li$_2$IrO$_3$ [Ref.
\onlinecite{cccmpaper}], with which \blio shares the same size
orthorhombic unit cell ($a \times b \times c$) and also many key
structural features. In both polytypes the iridium lattice is
locally three-fold coordinated and is made up of vertically-linked
zig-zag chains that alternate in orientation between the
$\bm{a}\pm\bm{b}$ direction. The only difference between them is
that in the hyperhoneycomb $\beta$-polytype this alternation
occurs between consecutive zig-zag chains, whereas in the
stripyhoneycomb $\gamma$-polytype the alternation occurs between
pairs of parallel zig-zag chains (which form a honeycomb row). The
magnetic structures in both polytypes are incommensurate with the
same propagation vectors within experimental error,
$\bm{q}$$=$$(0.57(1),0,0)$, the moments are counter-rotating
between the sites of every nearest-neighbor bond, and the plane of
rotation alternates between two orientations tilted away (by an
angle $\phi$) from the $ac$ plane between consecutive zig-zag
iridium chains vertically-linked along the $c$-axis. The $\phi$
angle was found to be somewhat smaller in $\gamma$-Li$_2$IrO$_3$,
but apart from this difference the magnetic structure in \glio can
be regarded as ``equivalent" to that in $\beta$-Li$_2$IrO$_3$ but
in a different lattice setting (for a formal mapping see Appendix
\ref{app:equivalence}). Those similarities are strongly suggestive
that the defining features of the magnetic structure, namely
non-coplanarity, the counter-rotation on every nearest-neighbor
bond, and the direction and even the magnitude of the
incommensurate ordering wavevector, are determined by the same
type of short-range magnetic interactions in both polytypes, and
those features appear to be robust against changes in the global
lattice connectivity (i.e. how zig-zag chains are
vertically-linked in a long-range pattern) which leave the local
connectivity of the zig-zag chains unchanged. In the
$\gamma$-polytype it was shown\cite{cccmpaper} that the key
features of the magnetic order can be stabilized by a spin
Hamiltonian with dominant Kitaev couplings with some additional
interactions; the observation of similar magnetic structures in
the $\beta$- and $\gamma$-polytypes suggests the same underlying
magnetic interactions occur in both cases. The relevant parent
Kitaev model for the $\beta$-polytype has been considered in
Refs.\ \onlinecite{lee,kimchi} and is illustrated in Fig.\
\ref{fig:magstruct}a) (unit cell 3) where the color of the bonds
indicates the anisotropy axis of the exchange, i.e. a
ferromagnetic Ising coupling between the components normal to the
Ir-O$_2$-Ir plane of each Ir-Ir bond, those directions define the
cubic axes $\mathsf{x}$, $\mathsf{y}$ and $\mathsf{z}$ related to
the crystallographic axes by $\mathsfbf{\hat{x}}
=(\bm{\hat{a}}+\bm{\hat{c}})/\sqrt{2}$, $\mathsfbf{\hat{y}}
=(\bm{\hat{a}}-\bm{\hat{c}})/\sqrt{2}$ and $\mathsfbf{\hat{z}}
=\bm{\hat{b}}$ (where we have assumed the parent ``idealized"
lattice with cubic IrO$_6$ octahedra and
$a$:$b$:$c$=1:$\sqrt{2}$:3). We note that the magnetic
interactions may also be very similar in magnitude between the
$\beta$ and $\gamma$-polytypes, as both materials show very
similar values for the magnetic ordering temperature.

\section{Conclusions}
\label{sec:conclusions} We have reported a combined study using
magnetic neutron powder diffraction and single-crystal magnetic
resonant x-ray diffraction experiments at the L$_3$ edge of Ir to
explore the magnetic structure of $\beta$-Li$_2$IrO$_3$, which is
a unique realization of a hyperhoneycomb iridium lattice, a
candidate to display 3D Kitaev physics of strongly-frustrated,
bond-anisotropic interactions. Using the azimuth dependence of the
magnetic Bragg peak intensities combined with magnetic symmetry
analysis we have obtained a complete solution for the magnetic
structure for all 16 iridium sites in the structural unit cell. We
find an incommensurate, non-coplanar magnetic structure with
moments counter-rotating on every nearest-neighbor bond. The
magnetic structure shows striking similarities to the magnetic
structure of the related structural polytype
$\gamma$-Li$_2$IrO$_3$, suggesting that the same underlying
Hamiltonian (with dominant Kitaev interactions) stabilizes the
defining features of the magnetic structure in both structural
polytypes.\\

\section{Acknowledgements}
\label{sec:acknowledgements} Work at Oxford was supported by EPSRC
(UK) and at Diamond and ISIS by STFC (UK). Work at Augsburg was
supported by the Helmholtz Virtual Institute 521 (``New states of
matter and their excitations").

\appendix
\section{Sample synthesis, characterization and low-temperature structural refinement}
\label{app:structure}

\begin{figure}[htbp]
\includegraphics[width=0.48\textwidth]{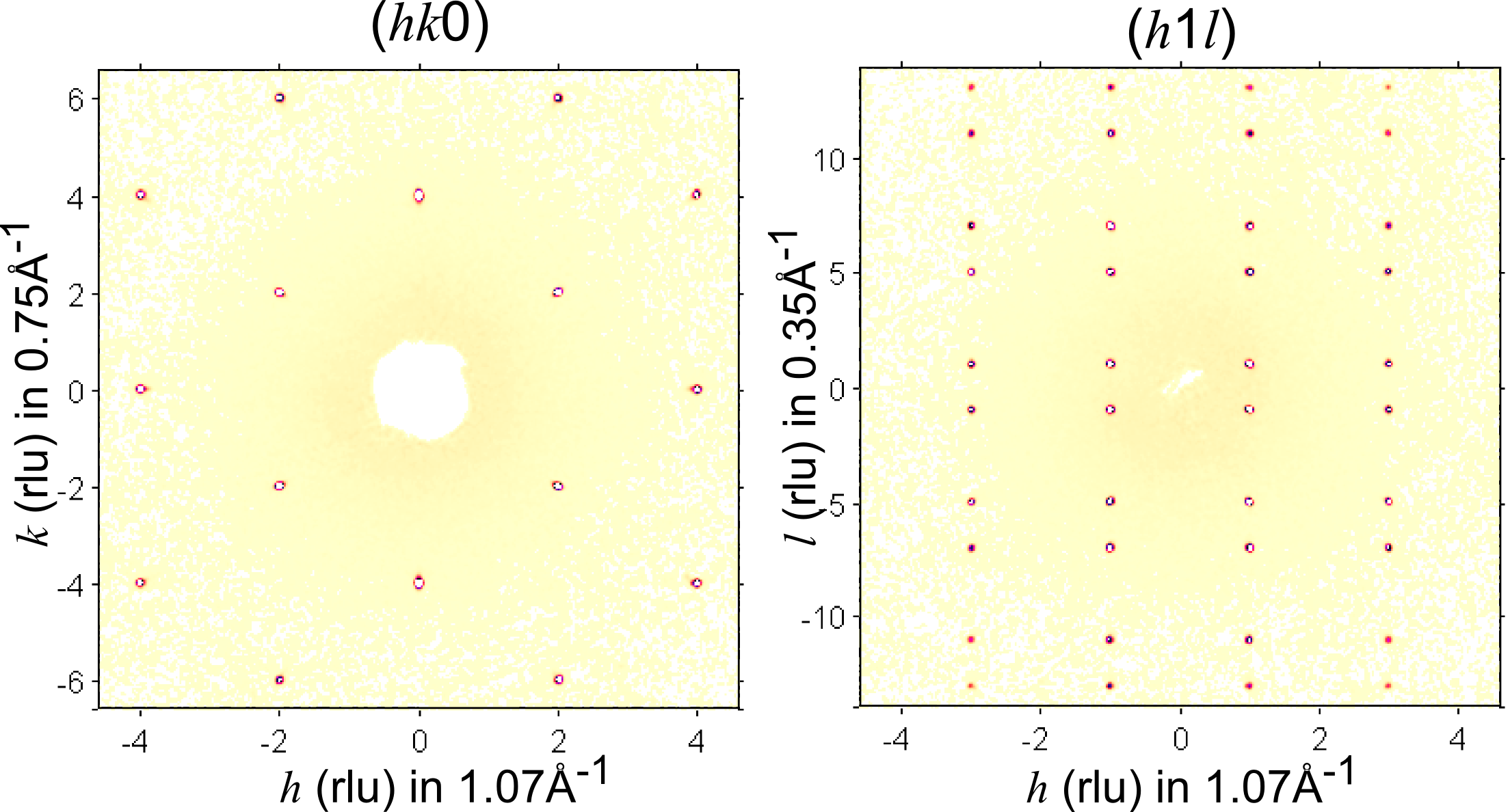}
\caption[]{(color online) X-ray diffraction pattern at 300\, K in
the ($hk0$) and ($h1l$) planes from the crystal used in the
resonant x-ray diffraction experiments.}
\label{fig:hk0_h1l_planes}
\end{figure}

The synthesis route to obtain the samples used in the neutron and
x-ray experiments was as follows. First, a powder sample of the
layered honeycomb phase $\alpha$-Li$_2$IrO$_3$ was prepared as
described in Ref.\ \onlinecite{singh-manni}. Then repetitive
annealing at 1100$^{\circ}$C transformed the sample into
hyperhoneycomb $\beta$-Li$_2$IrO$_3$, confirmed via powder x-ray
diffraction. Bulk magnetic susceptibility and heat capacity
measurements on the powder sample indicated evidence for magnetic
ordering at 38\, K, in full agreement with Ref.\
\onlinecite{takayama}. The polycrystalline powder contained small
shiny crystallites. Several such crystallites were extracted from
the powder and their diffraction pattern measured using a Mo
source single-crystal SuperNova x-ray diffractometer.
Representative diffraction patterns in the ($hk0$) and ($h1l$)
planes obtained from the sample used in the magnetic resonant
x-ray diffraction experiments (diameter 17$\mu$m) are shown in
Fig.\ \ref{fig:hk0_h1l_planes}. The patterns observe sharp Bragg
peaks, indicating a high-quality single crystal, with selection
rules consistent with the $Fddd$ space group expected for
$\beta$-Li$_2$IrO$_3$.
The results of the full structural refinement of a complete x-ray
diffraction data set collected at 100\, K from another crystal of
comparable dimensions from the same batch are given in Table\
\ref{struc_tab} where we have used the same convention for the
origin of the unit cell as in Ref.\ \onlinecite{takayama}. In the
table A.D.P. stands for Atomic Displacement Parameters, which were
assumed to be isotropic. The reliability factors of the Rietveld
refinement were $R(F^2)=4.50\%$, $wR(F^2)=5.24$\%, $R(F)=2.42$\%
and $\chi^2=0.318$.

\begin{table}[h!]
\caption{\label{struc_tab}\blio\ structural parameters at 100\, K}
\begin{ruledtabular}
\begin{tabular}{c c c c c}
\multicolumn{5}{l}{\textbf{Cell parameters}} \\
\multicolumn{5}{l}{Space group: $Fddd$ ($\#$70, origin choice 2)} \\
\multicolumn{5}{l}{Z = 16} \\
$a,b,c$ (\AA): & 5.8903(2) & 8.4261(3) & 17.7924(7) \\
$\alpha,\beta,\gamma$: & 90$^{\circ}$ & 90$^{\circ}$ & 90$^{\circ}$ \\
\multicolumn{5}{l}{Volume (\AA$^3$): 883.08(6)}\\
\\
\multicolumn{5}{l}{\textbf{Atomic fractional coordinates and isotropic A.D.P.'s}} \\
Atom & $x$ & $y$ & $z$ & $U_\mathrm{iso}$(\AA$^2$) \\
\hline
 Ir & 0.125 &  0.125 &  0.70845(7) &  0.0025(3) \\
 Li1 &  0.125 &  0.125 &  0.04167 &  0.00633 \\
 Li2 &  0.125 &  0.125 &  0.875 &  0.00633 \\
 O1 &  0.855(7) &  0.125 &  0.125 &  0.002(5) \\
 O2 &  0.621(8) &  0.3669(19) &  0.0384(7) &  0.002(3) \\

\\
\multicolumn{5}{l}{\textbf{Data collection}} \\
\multicolumn{5}{l}{$\#$ measured reflections: 3770} \\
\multicolumn{5}{l}{Data reduction $R_\mathrm{int}$: 6.99\%} \\
\multicolumn{5}{l}{(Criterion for observed reflections: $I > 2.0\sigma(I)$)} \\
\multicolumn{5}{l}{$\#$ observed independent reflections: 298} \\
\multicolumn{5}{l}{$\#$ fitted parameters: 9} \\
\end{tabular}
\end{ruledtabular}
\end{table}

\section{Magnetic Structure and Fourier Decomposition}
\label{app:magnetic}
This section lists the coordinates of the four iridium ions in the
primitive cell and also gives explicitly the expressions for the
magnetic moments at all sites in the unit cell and their Fourier
decomposition. The iridium ions occupy a single crystallographic
site with generating position ($1/8,1/8,z$) with $z=0.70445(7)$
and the four sites per primitive cell (labelled 1-4 in Fig.\
\ref{fig:magstruct}a) (left panel) have the coordinates listed in
Table.\ \ref{tab:magstruct}.

\begin{table}[tbh]
\caption{Fractional atomic coordinates of the iridium sites in the
primitive cell and corresponding magnetic basis vector components
in the determined magnetic structure.} \label{tab:magstruct}
\par
\begin{center}
\begin{tabular}
[c]{l|l|lll}\hline Site & Coordinates & $v_x$ & $v_y$ & $v_z$
\\\hline
 1 & $(0.125,0.125,z)$     & $+$ & $+$ & $+$\\
 2 & $(0.125,0.625,3/4-z)$ & $-$ & $+$ & $+$\\
 3 & $(0.375,0.375,1-z)$   & $-$ & $-$ & $+$\\
 4 & $(0.375,0.875,1/4+z)$ & $+$ & $-$ & $+$\\\hline
\end{tabular}
\end{center}
\end{table}

The determined magnetic structure is described by the basis vector
combination ($iA_x,iC_y,F_z$) with magnitudes $M_x$, $M_y$ and
$M_z$. The magnetic moment at position $\bm{r}$ belonging to site
index $n=1$$-$$4$ is given by
\begin{eqnarray}
\bm{M}_n(\bm{r}) & = & \bm{\hat{x}} M_x v_x(n) \sin
\bm{q}\cdot\bm{r} + \bm{\hat{y}} M_y v_y(n) \sin
\bm{q}\cdot\bm{r} \nonumber \\
&  + &\bm{\hat{z}} M_z v_z(n) \cos \bm{q}\cdot\bm{r},
\label{eq:Mr}
\end{eqnarray}
where $\bm{\hat{x}}$, $\bm{\hat{y}}$, $\bm{\hat{z}}$ are unit
vectors along the orthorhombic $\bm{a}$, $\bm{b}$, $\bm{c}$ axes,
respectively. The pre-factors $v_{x,y,z}$ are obtained from the
basis vectors along the corresponding axes, but without the
displacement phase factors $e^{-i\bm{q}\cdot(\bm{r}_n-\bm{r}_1)}$,
i.e. $v_x=[+--+]$ (from $A_x$), $v_y=[++--]$ (from $C_y$) and
$v_z=[++++]$ (from $F_z$). Eq.\ (\ref{eq:Mr}) describes all
iridium sites, including those related by $F$-centering
translations, where $\bm{r}$ is the actual position of the ion and
$n$ is the site index at the equivalent position (1$-$4) in the
primitive unit cell.

The magnetic moments are expressed in terms of their Fourier
components as $\bm{M}_n(\bm{r})=\sum_{\bm{k}=\pm\bm{q}}
\bm{M}_{\bm{k},n} e^{-i\bm{k}\cdot\bm{r}}$, where
\begin{eqnarray}
\bm{M}_{\bm{q},n} &=  & \left\{ i \left[\bm{\hat{x}} \frac{M_x}{2}
v_x(n)+ \bm{\hat{y}} \frac{M_y}{2} v_y(n) \right] \right. \nonumber \\
& + & \left. \bm{\hat{z}} \frac{M_z}{2} v_z(n) \right\}
e^{-i\bm{q}\cdot\left(\bm{r}_n-\bm{r}_1\right)}, \label{eq:Mq}
\end{eqnarray}
with $\bm{M}_{-\bm{q},n} =\bm{M}^*_{\bm{q},n}$ as the magnetic
moment distribution is real.\\

\section{Magnetic Structure Factors}
\label{app:magfactors}
This section gives the analytic expressions for the magnetic
structure factors for all the magnetic basis vectors. Starting
from the general expression in eq.\ (\ref{eq:structfact}) we
obtain the magnetic structure factors for the $C$, $A$, and $G$
basis vectors as follows
\begin{equation}
\bm{\mathcal{F}}^C(\bm{Q})= \left\{
\begin{array}{l}
16\bm{M}_{\pm \bm{q},1}\cos\frac{\pi l}{6} e^{i \xi_{ \pm}}, ~ h,k,l ~ \text{all even and}\\
\quad \quad \quad \quad \quad \quad h+k+l=4p+2, p ~\text{integer}\\
16\bm{M}_{\pm \bm{q},1}\cos\frac{\pi l}{6}\sin\frac{\pi (h+l-k)}{4} e^{i( \chi_{ \pm}+\pi/2)}, \\
\quad \quad \quad \quad \quad \quad   h,k,l  ~\text{all odd}\\
0, \quad \quad \quad \quad \quad ~ \text{otherwise}
\end{array} \right.
\label{eq:structC}
\end{equation}

\begin{equation}
\bm{\mathcal{F}}^A(\bm{Q})= \left\{
\begin{array}{l}
16\bm{M}_{\pm \bm{q},1}\sin\frac{\pi l}{6} e^{i (\xi_{ \pm}-\pi/2)}, ~ h,k,l ~ \text{all even and}\\
\quad \quad \quad \quad \quad \quad h+k+l=4p, p ~ \text{integer} \\
16\bm{M}_{\pm \bm{q},1}\sin\frac{\pi l}{6}\cos\frac{\pi (h+l-k)}{4} e^{i (\chi_{ \pm}-\pi/2)}, \\
\quad \quad \quad \quad \quad \quad   h,k,l  ~\text{all odd}\\
0, \quad \quad \quad \quad \quad ~ \text{otherwise}
\end{array} \right.
\label{eq:structA}
\end{equation}

\begin{equation}
\bm{\mathcal{F}}^G(\bm{Q})= \left\{
\begin{array}{l}
16\bm{M}_{\pm \bm{q},1}\sin\frac{\pi l}{6} e^{i (\xi_{ \pm}+\pi/2)}, ~ h,k,l ~ \text{all even and}\\
\quad \quad \quad \quad \quad \quad h+k+l=4p+2, p ~ \text{integer} \\
16\bm{M}_{\pm \bm{q},1}\sin\frac{\pi l}{6} \sin\frac{\pi (h+l-k)}{4}e^{i(\chi_{ \pm}+\pi)}, \\
\quad \quad \quad \quad \quad \quad   h,k,l  ~\text{all odd}\\
0, \quad \quad \quad \quad \quad ~ \text{otherwise},
\end{array} \right.
\label{eq:structG}
\end{equation}
where to obtain an analytic closed-form we used the ``ideal"
iridium positions $z$=5/8+1/12.

\section{Azimuth Dependence of the Magnetic Resonant X-ray Diffraction Intensity}
\label{app:MRXD}
\begin{figure}[htbp]
\begin{center}
\includegraphics[width=0.48\textwidth]{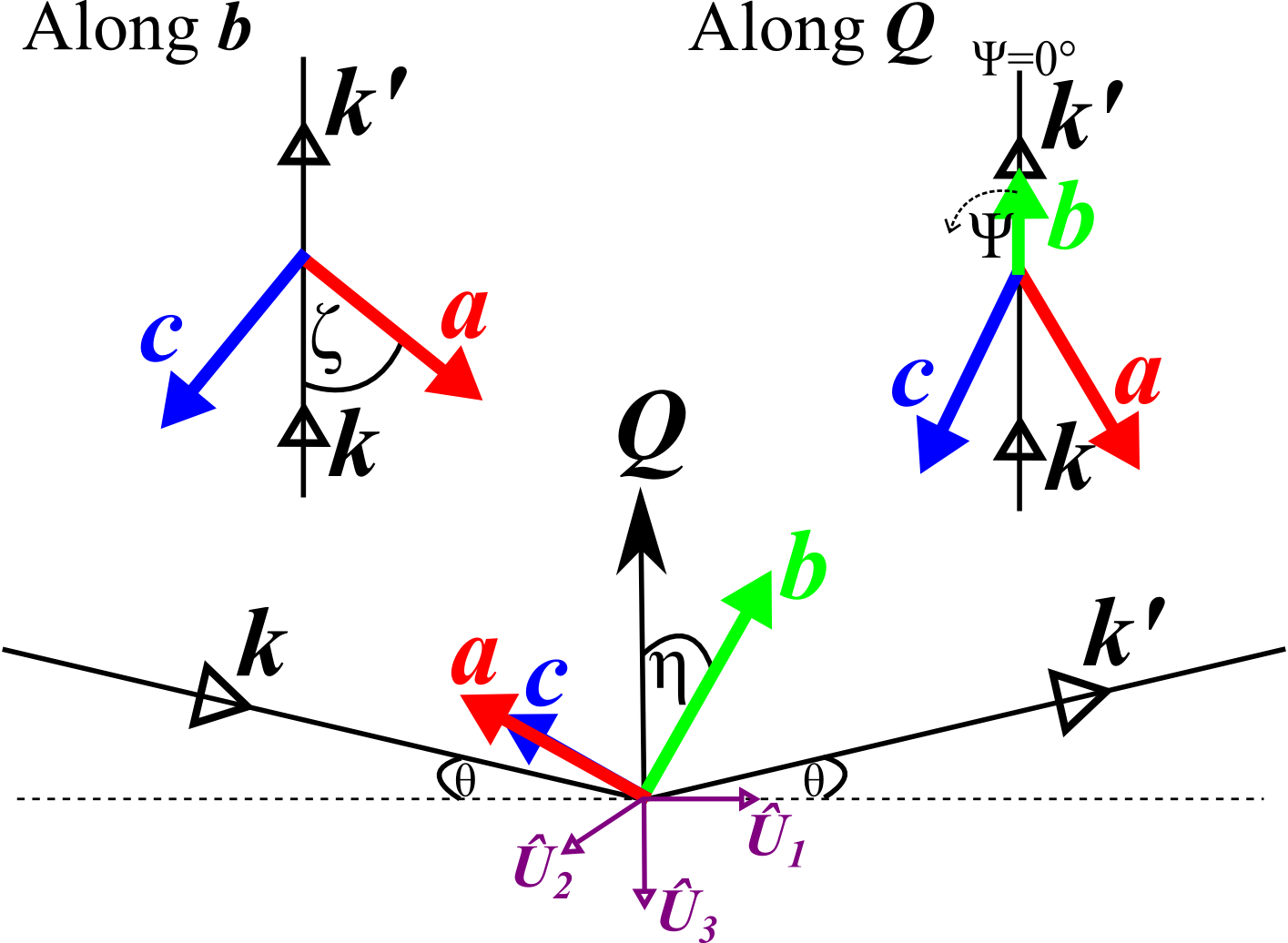}
\caption[]{Schematic diagram of the x-ray diffraction experiment
at zero azimuth ($\Psi=0$). $\bm{k}$ and $\bm{k'}$ are the
incident and scattered x-ray beam wavevectors, and
$\bm{Q}=\bm{k'}-\bm{k}$ is the scattering wavevector. Main
(bottom) panel shows the orientation of the Laboratory frame unit
vectors $\bm{\hat{U}_{1,2,3}}$ and the orthorhombic crystal axes
($\bm{a}$,$\bm{b}$,$\bm{c}$). Top left inset: view along $\bm{b}$
showing the angle $\zeta$ subtended from the scattering plane by
the $\bm{a}$ vector. Right inset: view along $\bm{Q}$. During an
azimuth scan the crystal axes are rotated around $\bm{Q}$ by an
angle $\Psi$ (in the sense show in the figure), the reference
position $\Psi=0$ is defined as the azimuth angle for which
$\bm{b}$ lies in the scattering plane making the smallest angle
with the incident beam direction
$\bm{\hat{k}}$.}\label{fig:expsketch}
\end{center}
\end{figure}

Here we obtain the azimuth dependence of the magnetic x-ray
diffraction intensity and show how it can be used to determine the
relative phase between magnetic basis vectors. As discussed in
Sec.\ \ref{sec:azimuths} the magnetic diffraction intensity is
proportional to $|\mathcal{F}_{\parallel}|^2$, where
$\mathcal{F}_{\parallel}= \bm{\mathcal{F}}\cdot \bm{\hat{k'}}$ is
the projection of the structure factor vector onto the scattered
beam direction. To calculate this dot product it is convenient to
express both vectors in terms of their components in the
Laboratory reference frame\cite{hill} illustrated in Fig.\
\ref{fig:expsketch} and defined by the orthonormal set of vectors
$\bm{\hat{U}_2}=\bm{\hat{k}} \bm{\times} \bm{\hat{k'}}$,
$\bm{\hat{U}_3}=-\bm{\hat{Q}}$ and
$\bm{\hat{U}_1}=\bm{\hat{U}_2}\bm{\times}\bm{\hat{U}_3}$, i.e.
$\bm{\hat{U}_2}$ is normal to the scattering plane and
$\bm{\hat{U}_1}$ is in the scattering plane making an angle
$\theta$ with both $\bm{k}$ and $\bm{k'}$, where $2\theta$ is the
total scattering angle. The scattered beam direction then can be
written as $\bm{\hat{k'}}=(\cos \theta, 0, -\sin \theta)_{L}$,
where the $L$ subscript indicates that the components are with
respect to the Laboratory basis of $\bm{\hat{U}_{1,2,3}}$ vectors,
as opposed to the Crystal basis, defined by the orthonormal set of
vectors $\bm{\hat{x}}$, $\bm{\hat{y}}$, $\bm{\hat{z}}$ (along the
orthorhombic axes $\bm{a}$, $\bm{b}$, $\bm{c}$).

The magnetic structure factor vector $\bm{\mathcal{F}}$ defined in
(\ref{eq:structfact}) is most naturally expressed in terms of the
Crystal basis as
($\mathcal{F}_x,\mathcal{F}_y,\mathcal{F}_z$)$_C$. The components
in the Laboratory basis
($\mathcal{F}_1,\mathcal{F}_2,\mathcal{F}_3$)$_L$ are obtained via
a linear transformation encoded in a matrix $T(\Psi)$ via
\begin{equation} \left[
\begin{array}{c}
\mathcal{F}_1\\
\mathcal{F}_2\\
\mathcal{F}_3
\end{array} \right]= T(\Psi) \left[
\begin{array}{c}
\mathcal{F}_x\\
\mathcal{F}_y\\
\mathcal{F}_z
\end{array} \right],
\label{eq:transformation}
\end{equation}
where we have explicitly introduced the dependence of the
transformation matrix on the azimuth angle $\Psi$, which defines
the rotation angle of the crystal axes around the scattering
wavevector $\bm{Q}$. To find the transformation matrix we first
consider the reference position of zero azimuth, $\Psi=0$,
depicted in Fig.\ \ref{fig:expsketch}. In this case the
orientation of the crystal axes with respect to the Laboratory
axes is specified by two angles $\eta$ and $\zeta$. $\eta$ is the
angle between the $\bm{b}$-axis and the scattering wavevector
$\bm{Q}$, given by
\begin{equation}
\eta=\cos^{-1} \bm{\hat{Q}}\cdot \bm{\hat{b}} \nonumber
\end{equation}
and $\zeta$ is the angle between the $\bm{a}$-axis and the
scattering plane, given by
\begin{equation}
\zeta =\text{tan}^{-1}\frac{\bm{Q} \cdot \bm{\hat{c}}}{\bm{Q}
\cdot \bm{\hat{a}}}~. \nonumber
\end{equation}
Using the diagram in Fig.\ \ref{fig:expsketch} the transformation
matrix is obtained as
\begin{equation} \label{eq:T0}
T(0)= \left [ \begin{array}{ccc}
-\cos \zeta \cos \eta & \sin \eta & -\sin \zeta \cos \eta \\
\sin \zeta & 0 & -\cos \zeta \\
- \cos \zeta \sin \eta & -\cos \eta & -\sin \zeta \sin \eta \\
\end{array} \right].
\end{equation}
When moving away from the reference position to a finite azimuth
angle $\Psi$ the coordinates in the Laboratory frame need to be
multiplied by the rotation matrix
\begin{equation} \label{eq:Rmat}
R(\Psi)= \left [ \begin{array}{ccc}
\cos \Psi & \sin \Psi & 0 \\
-\sin \Psi & \cos \Psi & 0 \\
0 & 0 & 1 \\
\end{array} \right],
\end{equation}
such that the resulting transformation matrix from the Crystal to
the Laboratory bases in (\ref{eq:transformation}) becomes
\begin{equation}
T(\Psi)=R(\Psi)~T(0). \label{eq:Tpsirough}
\end{equation}
Explicitly, substituting (\ref{eq:T0}) and (\ref{eq:Rmat}) into
(\ref{eq:Tpsirough}) yields
\begin{widetext}
\begin{equation} \label{eq:T}
T(\Psi)= \left [ \begin{array}{ccc}
\sin \Psi \sin \zeta -\cos \Psi \cos \zeta \cos \eta & \cos \Psi \sin \eta & -\cos \Psi \sin \zeta \cos \eta -\sin \Psi \cos \zeta\\
\sin \Psi \cos \zeta \cos \eta + \cos \Psi \sin \zeta & -\sin \Psi \sin \eta & \sin \Psi \sin \zeta \cos \eta -\cos \Psi \cos \zeta \\
- \cos \zeta \sin \eta & -\cos \eta & -\sin \zeta \sin \eta \\
\end{array} \right].
\end{equation}
\end{widetext}
Once the structure factor components are obtained in the
Laboratory basis, then the projection of the structure factor
vector along the scattered beam direction is readily obtained as
\begin{equation}\label{eq:structpara}
\mathcal{F}_{\parallel}=\cos \theta~ \mathcal{F}_1  - \sin \theta
~ \mathcal{F}_3.
\end{equation}

To illustrate the azimuth dependence of the magnetic x-ray
diffraction intensity we consider the magnetic peak at
$(-6,6,8)+\bm{q}$ in Fig.\ \ref{fig:phaseazi}a). Here only the
$A_x$ and $F_z$ basis vectors contribute, therefore
$\bm{\mathcal{F}}=(\mathcal{F}_x,0,\mathcal{F}_z$)$_C$ where the
structure factor components $\mathcal{F}_x$ and $\mathcal{F}_z$
are of the type $\mathcal{F}^A$ and $\mathcal{F}^F$ in
(\ref{eq:structA},\ref{eq:structF}), respectively. Using
(\ref{eq:structpara}) combined with (\ref{eq:transformation}) and
(\ref{eq:T}) we obtain
\begin{equation}
\mathcal{F}_{\parallel}(\Psi)=g_x(\Psi)~\mathcal{F}_x +
g_z(\Psi)~\mathcal{F}_z, \nonumber
\end{equation}
where the geometric factors are
\begin{eqnarray}
g_x(\Psi) & = &(\sin\Psi \sin\zeta-\cos\Psi\cos\zeta\cos\eta) \cos
\theta + \nonumber \\
  & & \cos \zeta \sin
\eta \sin \theta  \nonumber \\
g_z(\Psi) & = & \sin\zeta \sin\eta \sin \theta - \nonumber \\
& & (\cos\Psi \sin\zeta\cos\eta + \sin \Psi \cos \zeta) \cos
\theta. \nonumber \label{eq:structparafddd}
\end{eqnarray}
The magnetic scattering intensity is then proportional to
\begin{align}
|\mathcal{F}_{\parallel}(\Psi)|^2&=|g_x(\Psi)~\mathcal{F}_x|^2 + |g_z(\Psi)~{\mathcal{F}}_z|^2 \nonumber \\
& {} +2{\rm{Re}}(\mathcal{F}_z\mathcal{F}_x^*)~g_x(\Psi)
g_z(\Psi), \nonumber \label{eq:structfull}
\end{align}
where ${\rm Re}()$ means the real part. The first two terms give
the sum of the separate contributions from the two basis vectors,
whereas the last (cross-term) is sensitive to the relative phases
between the two basis vectors. Using eqs.\
(\ref{eq:structF},\ref{eq:structA}) the cross-term is proportional
to
\begin{equation}
\mathcal{F}_z\mathcal{F}_x^* \propto \left\{
\begin{array}{l l}
\mp i\sin\frac{\pi l}{6}\cos\frac{\pi l}{6} & ,\quad F_z \pm
A_x\\
\pm\sin\frac{\pi l}{6}\cos\frac{\pi l}{6} & ,\quad F_z \pm iA_x.
\end{array} \right.
\label{eq:structxz}\nonumber
\end{equation}
The cross-term is therefore directly sensitive to the phase
difference between the two contributing basis vectors. If $F_z$
and $A_x$ are in-phase or $\pi$ out-of-phase (denoted as $F_z \pm
A_x$) the product $\mathcal{F}_z\mathcal{F}_x^*$ is purely
imaginary, so the cross-term cancels and the intensity is given by
the sum of the contributions from the two basis vectors taken
separately. This is clearly inconsistent with the data (purple
curve in Fig.\ \ref{fig:phaseazi}a)). A phase difference of $\pm
\pi/2$ between the two basis vectors (denoted as $F_z \pm iA_x$)
gives a finite cross-term, which changes sign between those two
cases, and the data in Fig.\ \ref{fig:phaseazi}a) identifies the
basis vector combination as $F_z + iA_x$ (red solid line).
\begin{table}[htb]
\caption{\label{table_equivalence_sites} Equivalence between the
iridium sites in \glio ($Cccm$ space group) with those in \blio
($Fddd$ space group) (when both are projected onto the $ac$
plane). Last column shows additional translations to be applied to
primitive lattice sites in the $Fddd$ structure (column 2) in
addition to a common translation by (-1/8,0,1/8) to get the
positions in the $Cccm$ structure (column 1). For each structure
the ``ideal" iridium positions have been used.}
\begin{tabular}{c|c|c}
Site &  Site  & Translation \\
$Cccm$ &  $Fddd$ & \\
\hline
1 & 4 & -(0,0,1) \\
2 & 3 &  (0,0,0)\\
3 & 3 & (0,1/2,1/2) \\
4 & 4 & -(0,1/2,1/2) \\
\hline
1$'$ & 2 & (1/2,1/2,0) \\
2$'$ & 1 & (1/2,0,-1/2) \\
3$'$ & 1 & (1/2,1/2,0) \\
4$'$ & 2 & (1/2,1,1/2) \\
\end{tabular}
\end{table}
\section{Mapping of magnetic basis vectors between $\bm{\beta}$- and
$\bm{\gamma}$-L\MakeLowercase{i}$_2$I\MakeLowercase{r}O$_3$}
\label{app:equivalence}

In this section we describe a formal mapping of the magnetic basis
vectors between the notation used for the $\beta$- (space group
$Fddd$) and $\gamma$-polytypes of Li$_2$IrO$_3$ (space group
$Cccm$). The iridium lattice arrangement in both crystal
structures is made up of zig-zag chains vertically-linked along
the $c$-axis, with the only difference being that in the
$\beta$-polytype successive chains along $c$ alternate in
orientation between the $\bm{a}\pm\bm{b}$ directions, whereas in
the $\gamma$-polytype a pair of successive chains are parallel and
this direction then alternates for the next pair of parallel
chains and so on, between the $\bm{a}\pm\bm{b}$ directions. When
both structures are viewed in projection onto the $ac$ plane [as
shown in Fig.\ \ref{fig:magstruct}a) and Fig.\ 4 of Ref.\
\onlinecite{cccmpaper}] the two iridium lattice arrangements then
look identical. This suggests that a formal mapping can be
constructed between the notations used to describe the sites of
the primitive cells of the two materials, and also between the
magnetic basis vectors, such that one can then describe a magnetic
structure equivalently in the notation of one or the other space
group (of course the magnetic structures would only be
``equivalent" up to the projection of the sites onto the
$b$-axis). Table\ \ref{table_equivalence_sites} lists the
equivalence between the sites in the primitive cell of the two
structures (up to the $y$-coordinate of sites), where for the
$\gamma$-polytype we follow the notation in Ref.\
\onlinecite{cccmpaper} (two iridium sublattices with sites 1-4 and
1$'$-4$'$). Using this site equivalence one can construct a
mapping of the magnetic basis vectors as listed in Table\
\ref{table_equivalence_vectors}.
\begin{table}[htb]
\caption{\label{table_equivalence_vectors} Mapping of magnetic
basis vectors between the $Fddd$ and $Cccm$ notations.}
\begin{tabular}{c|c}
Basis vector   & Basis vector  \\
$Fddd$ (4-site [1-4]) & $Cccm$ (8-site,[1-4,1$'$-4$'$])\\
\hline
$A$ [$+--+$] & ($A,-A$) [$+--+-++-$] \\
$C$ [$++--$] & ($-F,F$) [$----++++$]  \\
$F$ [$++++$] & ($F,F$) ~~ [$++++++++$] \\
\end{tabular}
\end{table}
The magnetic structure described by the basis vector combination
($iA_x,iC_y,F_z$) in the $Fddd$ setting then corresponds, in the
notation of the $Cccm$ setting, to $i(A,-A)_x,i(-1)^m(F,-F)_y,
(F,F)_z$ with $m=1$.
\end{document}